\documentclass[a4paper,fleqn ]{cas-sc}
\def\tsc#1{\csdef{#1}{\textsc{\lowercase{#1}}\xspace}}
\tsc{WGM}
\tsc{QE}
\usepackage[sort&compress,square,numbers]{natbib}

\usepackage{float}
\usepackage{booktabs}
\usepackage{graphicx}
\usepackage{adjustbox} 
\usepackage{caption}
\usepackage{siunitx}

\usepackage{subfigure}
\usepackage{threeparttable}
\usepackage[labelsep=period]{caption}
\usepackage[colorlinks]{hyperref}
\colorlet{scolor}{black}
\colorlet{hscolor}{DarkSlateGrey}
\usepackage{amsmath}
\usepackage[dvipsnames]{xcolor}
\definecolor{mypink1}{rgb}{048,104,141}
\hypersetup{colorlinks=true,linkcolor=Cerulean, filecolor=Cerulean,urlcolor=Cerulean,citecolor=Cerulean,CJKbookmarks=true} 
\begin{document}
	\captionsetup[figure]{labelfont={bf},labelsep=period,name={Fig. }}
	\let\printorcid\relax
	\let\WriteBookmarks\relax
	\def\floatpagepagefraction{1}
	\def\textpagefraction{.001}
	\shorttitle{}
	\title [mode = title]{A personalized model and optimization strategy for estimating blood glucose concentrations from sweat measurements}                      

\author[mymainaddress]{\textcolor[RGB]{0,0,1}{Xiaoyu Yin}}\ead{x.yin@tue.nl}
	\cormark[1]

\author[mymainaddress]{\textcolor[RGB]{0,0,1}{Elisabetta Peri}}

\author[mymainaddress]{\textcolor[RGB]{0,0,1}{Eduard Pelssers}}

\author[mymainaddress]{\textcolor[RGB]{0,0,1}{Jaap den Toonder}}

\author[mysecondaryaddress]{\textcolor[RGB]{0,0,1}{Lisa Klous}}

\author[mythirdaddress]{\textcolor[RGB]{0,0,1}{Hein Daanen}}

\author[mymainaddress]{\textcolor[RGB]{0,0,1}{Massimo Mischi}}
	
\address[mymainaddress]{Eindhoven University of Technology, Eindhoven, NL}

\address[mysecondaryaddress]{Netherlands Organisation for Applied Scientific Research, Soesterberg, NL}

\address[mythirdaddress]{Vrije Universiteit Amsterdam, Amsterdam, NL}

\begin{abstract}
Background and objective: Diabetes is one of the four leading causes of death worldwide, necessitating daily blood glucose monitoring. While sweat offers a promising non-invasive alternative for glucose monitoring, its application remains limited due to the low to moderate correlation between sweat and blood glucose concentrations, which has been obtained until now by assuming a linear relationship. This study proposes a novel model-based strategy to estimate blood glucose concentrations from sweat samples, setting the stage for non-invasive glucose monitoring through sweat-sensing technology. 

Methods: We first developed a pharmacokinetic glucose transport model that describes the glucose transport from blood to sweat. Secondly, we designed a novel optimization strategy leveraging the proposed model to solve the inverse problem and infer blood glucose levels from measured glucose concentrations in sweat. To this end, the pharmacokinetic model parameters with the highest sensitivity were also optimized so as to achieve a personalized estimation. Our strategy was tested on a dataset composed of 108 samples from healthy volunteers and diabetic patients.

Results: Our glucose transport model improves over the state-of-the-art in estimating sweat glucose concentrations from blood levels (higher accuracy, p$<$0.001). Additionally, our optimization strategy effectively solved the inverse problem, yielding a Pearson correlation coefficient of 0.98 across all 108 data points, with an average root-mean-square-percent-error of 12$\% \pm$8$\%$. This significantly outperforms the best sweat-blood glucose correlation reported in the existing literature (0.75). 

Conclusion: Our innovative optimization strategy, also leveraging more accurate modeling, shows promising results, paving the way for non-invasive blood glucose monitoring and, possibly, improved diabetes management. 

\end{abstract}


\maketitle  
\section{Introduction}

 Diabetes is a chronic condition characterized by elevated blood glucose concentrations, caused by either insufficient insulin production or the body's ineffective use of insulin. Diabetes is emerging as a major global threat worldwide \citep{bloom2018carga1}. The prevalence of diabetes was estimated to be around 537 million people globally in 2021 \citep{DE2}, which is expected to rise to 643 million by 2030 and further to 783 million by 2045 \citep{DE2,saeedi2019global3}. Diabetes not only affects a large portion of the global population but is also a leading driver of mortality, ranking among the top 10 causes of death in adults \citep{DE2,saeedi2019global3}. In 2021 alone, approximately 6.7 million adults between the ages of 20 and 79 died from diabetes or associated complications, accounting for 12.2\% of all deaths in this age group \citep{DE2}. The economic impact of diabetes is substantial, with direct health expenditures nearing one trillion USD in 2021 \citep{DE2}. This multifaceted impact of diabetes demonstrates the urgent need for the development of innovative strategies to improve patient management and outcomes. Monitoring of diabetes is vital as it allows for the adjustment of insulin doses in response to blood glucose concentrations and facilitates the adoption of a more appropriate diet \citep{hermanns2022coordination4}. The medical standards of the American Diabetes Association recommend that diabetic patients receiving regulated pharmacological treatment perform self-monitoring of blood glucose 6-8 times daily to detect and prevent asymptomatic hypoglycemia and hyperglycemia \citep{grant2015standards5}. 

The most common method for daily glucose monitoring is fingerstick, which involves using a lancet to prick the fingertip and obtain a small blood sample for rapid tests \citep{hoffman2023minimally6}. Additionally, there are other monitoring methods, such as those based on microneedles for transdermal sampling, which enable glucose monitoring through the interstitial fluid \citep{lu2022microneedle7, Leelarathna2022}. Nevertheless, both blood and interstitial fluid sampling techniques are invasive. Invasive sampling may cause discomfort and inconvenience to the subject. Additionally, repeated sampling will amplify these effects, thereby limiting the frequency at which invasive sampling can be performed and potentially reducing patient compliance. 

With the development of novel non-invasive sweat-sensing technology, semi-continuous monitoring of sweat glucose concentrations will become achievable in the near future. This innovation utilizes autonomous sweat-sensing patches that require minimal effort from patients at home and eliminate the need for blood sampling \citep{Hermans2020,Moonen2020,VanEnter2018}. This approach not only enables regular updates on an individual's glucose status, but also avoids the drawbacks of invasiveness associated with traditional blood sampling methods. Consequently, sweat-sensing techniques could serve as a possible solution for prolonged and less invasive glucose monitoring \citep{Davis2024}, but this is only feasible if the relationship between sweat and blood glucose concentrations is confirmed. 

While several studies have focused on the development of glucose sweat sensors, only six studies have explored the correlation between glucose concentrations in sweat and blood \citep{moyer2012correlation8,Klous2021,Nyein2019,Karpova2019,Mwaurah2024,Sempionatto2021}. The results obtained so far show only low to moderate correlation, limiting the potential clinical application of sweat-based glucose monitoring. Four of these studies utilized linear regression methods for the analysis \citep{moyer2012correlation8,Klous2021,Nyein2019,Karpova2019}. Specifically, Moyer and colleagues demonstrated a moderate correlation coefficient of 0.75 from a total of 115 experimental samples, which included 7 diabetic patients \citep{moyer2012correlation8}. Similarly, Klous et al. reported a correlation coefficient of 0.73 from 48 data points gathered from 12 healthy individuals \citep{Klous2021}. Nyein et al. analyzed 48 data points from 28 diabetic patients and 20 healthy individuals, finding a low correlation coefficient of 0.3 \citep{Nyein2019}. Karpova et al. reported a correlation coefficient of 0.75 from 25 data points, which included 19 healthy participants \citep{Karpova2019}. The remaining two studies \citep{Mwaurah2024,Sempionatto2021} were excluded from detailed analysis due to their very small sample size ($\leq 3$ participants), making their findings less robust and generalizable for wide-spread application. These findings suggest that the assumed linear relationship may not adequately capture the physiological dynamics governing glucose transport. 

Recognizing this limitation, La Count and colleagues proposed a pharmacokinetic glucose transport model, taking into account the passive transport mechanisms of glucose, and used a dynamic mathematical modeling approach for the model simulations \citep{la2019modeling9}. This model was designed to predict sweat glucose concentrations based on measured glucose concentrations in blood, using 45 data points across five datasets from the literature. Nonetheless, the model proposed by La Count and colleagues did not consider the influence of varying sweating rates on the dilution of glucose concentrations, attributable to different water inflows entering the sweat glands under these conditions. Additionally, the values of the model parameters, such as the diffusion coefficient in water and the glucose uptake rate, were derived from existing literature, which is often based on small sample sizes or population averages \citep{mari2020mathematical10,longo1989increased11}. Since individuals may display varying physiological characteristics, their corresponding biophysical parameter values may also differ \citep{basu2006effects12,mendes2019assessment13}. As a result, inaccuracies could arise due to inter-individual variability when models use averaged, literature-derived, parameter values. Therefore, a new model is required that takes into account the dilution effect due to varying degrees of water inflow into the sweat glands, as well as for inter-individual variations in biophysical parameter values. 

Notably, La Count et al.'s model is specifically designed to estimate sweat glucose concentrations from blood glucose concentrations. However, the primary challenge for the clinical application of sweat-based glucose monitoring involves estimating blood glucose concentrations from sweat glucose concentrations, an inverse problem that remains unsolved in the literature \citep{Davis2024}. The inherent complexity of these pharmacokinetic models makes the analytical derivation of inverse problem solutions particularly challenging \citep{zenker2007inverse14}. To date, no studies have successfully addressed this inverse problem to enable estimation of blood glucose concentrations from sweat measurements, underscoring the need for innovative approaches in this area.

The purpose of this research is twofold. First, we propose a glucose transport model that describes how glucose is transported from the blood to sweat along a single sweat gland, detailed in Section 2.2. This model enables the prediction of sweat glucose concentrations based on variations in blood glucose concentrations. Secondly, we present an innovative double-loop optimization strategy, aimed at solving the inverse problem of estimating glucose concentrations in blood based on measured sweat glucose concentrations, detailed in Section 2.3. This approach builds upon and improves over our preliminary work \citep{yin2023estimation28}. The double-loop strategy incorporates two intertwined loops. The first loop aims at accurately estimating blood glucose concentrations for a given parameterization of the glucose transport model and the measured sweat glucose concentration, while the second loop optimizes the model parameters to closely align with the actual physiological values of individuals, thereby yielding more accurate patient-specific estimation. 

\section{Methods}\label{sec2}

\subsection{Dataset and Performance Measures}
In this study, we specifically selected datasets that report both sweat and blood glucose concentrations, along with corresponding sweat rates, which are critical parameters for our model. We used a total of seven experimental datasets comprising 108 measurement points from eighteen healthy volunteers and two diabetic patients. Six of these datasets were obtained from the literature \citep{moyer2012correlation8,la2019modeling9,lee2016graphene15,boysen1984modified16,Nyein2019}. All data used in this paper were fully anonymous, with no personal information included, as obtained from published papers. Hence, no additional ethical approval was required for their use. The seventh dataset was acquired from Vrije Universiteit Amsterdam, with the approval of the Medical Ethics Committee of Erasmus MC Rotterdam (MEC-2019-0202), in accordance with the revised Declaration of Helsinki (2013). This particular dataset involved twelve participants, all of whom were fully informed about the study procedures and potential risks before they provided their written informed consent \citep{Klous2021}. 

In Table~\ref{t1}, an overview of the included dataset is provided, noting the number of subjects per dataset, along with the methods used to induce and collect sweat, and the approaches employed to vary blood glucose concentrations. Blood and sweat glucose concentrations were acquired together with the corresponding sweat rate for each study. Further details of the experimental setup are reported in \citep{moyer2012correlation8,la2019modeling9,lee2016graphene15,boysen1984modified16,Nyein2019,Klous2021}. 

\begin{table*}
	\caption{Overview of the Utilized Datasets of Sweat and Blood Measurements}
	\label{t1}
    \centering
	\scalebox{1}{
	\begin{tabular}{lllllll}
		\hline
		\begin{tabular}[l]{@{}l@{}}Data\\ Source\end{tabular} &
		\begin{tabular}[l]{@{}l@{}} Subject Health \\ Condition\end{tabular} &
		\begin{tabular}[l]{@{}l@{}} Blood Glucose \\ Variation Method\end{tabular} &
		\begin{tabular}[l]{@{}l@{}}Sweat Collection\\ Method\end{tabular} &
		\begin{tabular}[l]{@{}l@{}}Perspiration\\Method\end{tabular} &
		\begin{tabular}[l]{@{}l@{}}Measurement \\ Points\end{tabular} \\ 
		\hline
		Exp 1 \cite{lee2016graphene15}& 1 Healthy & Meal intake & Wearable patch & \begin{tabular}[l]{@{}l@{}}Warm condition \\ sweating\end{tabular} & 15 \\
		
		Exp 2 \cite{boysen1984modified16} & 2 Healthy & \begin{tabular}[l]{@{}l@{}}Wet glucose \\ ingestion\end{tabular} & \begin{tabular}[l]{@{}l@{}}Polyethylene film \\ collector\end{tabular} & Sauna & 19 \\
	
		Exp 3 \cite{moyer2012correlation8} & 1 Diabetic & \begin{tabular}[l]{@{}l@{}}Wet glucose \\ ingestion\end{tabular} & \begin{tabular}[l]{@{}l@{}}Serpentine chamber \\ perfusion\end{tabular} & Iontophoresis & 6 \\
	
		Exp 4 \cite{la2019modeling9} & 1 Healthy & Fasting & \begin{tabular}[l]{@{}l@{}}Hydrophilic mesh \\ disks\end{tabular} & Iontophoresis & 7 \\
	
		Exp 5 \cite{la2019modeling9} & 1 Healthy & Fasting & \begin{tabular}[l]{@{}l@{}}Hydrophilic mesh \\ disks\end{tabular} & Iontophoresis & 5 \\ 

		Exp 6 \cite{Nyein2019} & \begin{tabular}[l]{@{}l@{}}1 Healthy and \\ 1 Diabetic\end{tabular} & Fasting & Wearable patch & Iontophoresis & 8 \\

		Exp 7 \cite{Klous2021} & 12 Healthy & \begin{tabular}[l]{@{}l@{}}No dietary \\ restrictions\end{tabular} & Absorbent patch & Cycling & 48 \\

		\hline
	\end{tabular}}
\end{table*}

\subsection{Glucose Transport Model}
\begin{figure*}[h]
\centering
\includegraphics[width=0.9\textwidth]{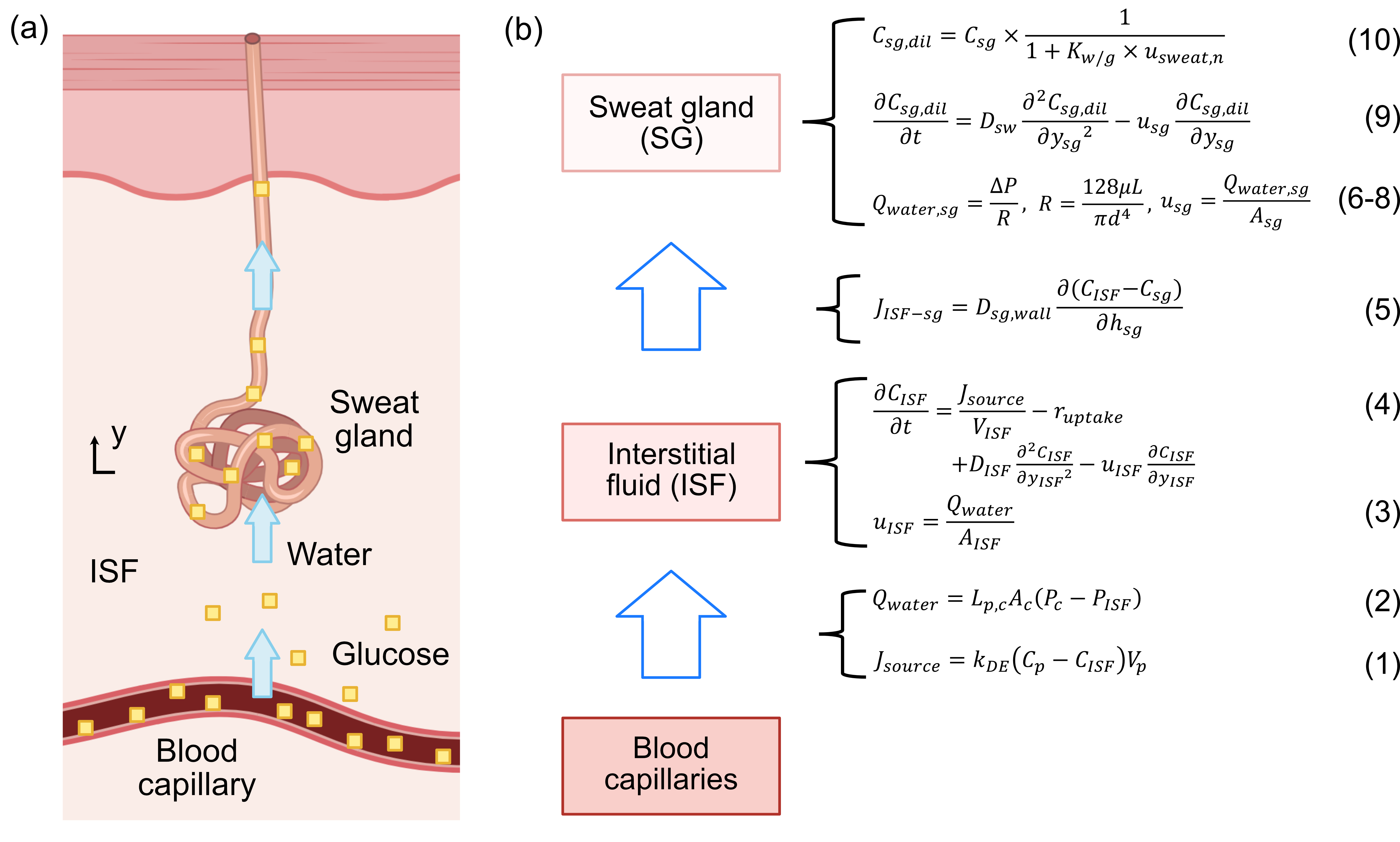}
\caption{(a) Schematic of the glucose transport process from blood to sweat along a single sweat gland. (b) Compartment model and corresponding formulas. }\label{fig1}
\end{figure*}

Our glucose transport model, which estimates glucose concentrations in sweat from measured glucose concentrations in blood, is based on passive transport mechanisms of glucose. This model is segmented into three compartments: the blood capillary, interstitial fluid (ISF), and sweat gland. Unlike the model developed by La Count et al. \citep{la2019modeling9}, which is the only existing framework in this field and includes 27 parameters, our model uses 18 parameters and clearly details the movement of glucose within the ISF and its transport mechanisms across the sweat gland wall from the ISF to the sweat gland. Additionally, our model integrates the dynamics of water movement from the blood capillaries through the sweat gland, ultimately forming sweat, and accounts for the dilution effect of varying sweat rates on glucose concentrations. The entire process is schematically illustrated in Fig.~\ref{fig1}.

\subsubsection{Glucose Transport Process}
Initially, glucose is transported from the blood capillary compartment into the ISF compartment, driven by the glucose concentration gradient between these two compartments, which is described as
\begin{equation}
    J_{source}=k_{DE}(C_p-C_{ISF})V_p,\label{equ1}
\end{equation}
where $J_{source}$ in $mol\ s^{-1}$ represents the glucose flow rate out of the blood capillary compartment, $C_p$ is the plasma glucose concentration in blood capillaries in $mol\, m^{-3}$, $C_{ISF}$ is the glucose concentration in the ISF in $mol\, m^{-3}$, $k_{DE}$ is the dermal clearance constant of glucose in $s^{-1}$ \citep{Ibrahim2012}, representing the rate at which glucose is drawn from the capillaries to the ISF, and $V_p$ represents the effective volume of the blood capillary compartment in $m^{3}$. The blood capillary compartment is modeled as a cylindrical segment with a radius of \SI{4}{\micro\meter} and a length of \SI{600}{\micro\meter} \citep{Himeno2016}, reflecting the dimensions of a capillary segment associated with a single sweat gland \citep{Wilke2007sweatgland}.

Simultaneously, water flows from the blood capillary compartment into the ISF compartment, driven by differences in hydrostatic pressure between the capillaries and the interstitial spaces. This process is described by the Starling equation as
\begin{equation}
    Q_{water} = L_{p,c} A_c  (P_c - P_{ISF}),\label{equation1}
\end{equation}
where $Q_{water}$ in $m^{3}\ s^{-1}$ represents the water flow rate out of the blood capillary compartment, $L_{p,c}$ represents the capillary hydraulic conductivity of water in $m\ s^{-1} \ mmHg^{-1}$ \citep{Kellen2003}, $A_c$ is the effective surface area of the blood capillary compartment in $m^{2}$ \citep{Himeno2016}, $P_c$ and $P_{ISF}$ are the capillary and interstitial hydrostatic pressure in $mmHg$, respectively \citep{Haggerty2019}. Since the capillary walls are completely permeable to small molecules like water, the influence of the colloid osmotic pressure gradient is generally neglected \citep{Rippe1986}.

The velocity of water flow in the ISF compartment, denoted as $u_{ISF}$ in $m\,s^{-1}$, can be expressed as the ratio of the water flow rate $Q_{water}$ to the effective cross-sectional area of the ISF compartment $A_{ISF}$ in $m^{2}$ \citep{Himeno2016}, and is given as
\begin{equation}
u_{ISF} = \frac{Q_{water}}{A_{ISF}}. \label{equation2}
\end{equation}
By combining (\ref{equation1}) and (\ref{equation2}), the derived formula demonstrates that the velocity of water flow in the ISF compartment $u_{ISF}$ is directly influenced by the ratio of $A_{c}/A_{ISF}$, namely, the relative areas of the blood capillary and the ISF compartments. 

After glucose flows into the ISF compartment, glucose is partially absorbed due to intracellular metabolism, and undergoes diffusion and convection processes. These processes are collectively described as
\begin{equation}
    \frac{\partial C_{ISF}}{\partial t} = \frac{J_{source}}{V_{ISF}} - r_{uptake} + D_{ISF} \frac{\partial^2 C_{ISF}}{\partial y_{ISF}^2} - u_{ISF} \frac{\partial C_{ISF}}{\partial y_{ISF}},\label{equ2}
\end{equation}
where $r_{uptake}$ represents the rate of glucose uptake due to cellular metabolism in $mol\ m^{-3}\,s^{-1}$ \citep{longo1992glucose17}, $y_{ISF}$ in $m$ represents the spatial displacement of glucose from its entry point of the ISF compartment, $D_{ISF}$ is the diffusion coefficient of glucose in ISF in $m^2\,s^{-1}$ \citep{Khalil2006}, and $V_{ISF}$ is the effective volume of the ISF compartment in $m^{3}$. The ISF compartment is modeled based on the Krogh cylinder model, which assumes that each capillary is surrounded by a cylinder of ISF. The radius of the ISF cylinder is set at \SI{18}{\micro\meter}, and the length of the ISF cylinder is set to be identical to that of the blood capillary compartment \citep{Himeno2016}. Consequently, when (\ref{equ1}) is substituted into (\ref{equ2}), one can observe that the ratio $V_{p}/V_{ISF}$ effectively cancels out the identical lengths of the blood capillary and ISF compartments. Thus, the length of these two compartments does not influence the subsequent calculations.

Subsequently, glucose moves from the ISF compartment into the sweat gland compartment, driven by the concentration gradient between these two compartments. The transport of glucose across the sweat gland wall is dominated by diffusion. The impact of convection during this process is then considered negligible \citep{Lavery2001}. This process can be described using the Fick's first law as
\begin{equation}
    J_{ISF-sg} = D_{sg,wall} \frac{\partial (C_{ISF} - C_{sg})}{\partial h_{sg}},\label{equ3}
\end{equation}
where $J_{ISF-sg}$ in $mol\ s^{-1}$ represents the glucose flux that enters the sweat gland compartment from the ISF compartment, $D_{sg,wall}$ is the diffusion coefficient of glucose for the sweat gland wall in $m^2\,s^{-1}$ \citep{zhang2005effective21}, $C_{sg}$ represents the glucose concentration in sweat gland in $mol\ m^{-3}$, and $h_{sg}$ represents the thickness of the sweat gland wall in $m$ \citep{Sonner2015}. 

Water also flows from the ISF compartment to the sweat gland compartment, subsequently moving along the sweat gland to the surface of the human skin. This process is driven by a pressure difference, caused by an increase in interstitial pressure due to water inflow from the blood capillary compartment into the ISF compartment \citep{Aronson2022}, and can be described using Darcy's law as

\begin{equation}
    Q_{water,sg} = \frac{\Delta P}{R},\label{equation3}
\end{equation}
where $Q_{water,sg}$ in $m^{3}\ s^{-1}$ represents the water flow rate inside the sweat gland, $\Delta P$ represents the pressure difference across the sweat gland in $mmHg$ \citep{Schulz1969}, $R$ is the hydraulic resistance of the sweat gland in $Pa\cdot s\ m^{-3}$ \citep{Sonner2015}, which can be expressed as

\begin{equation}
    R = \frac{128 \mu L}{\pi d^4},\label{equation4}
\end{equation}
where $\mu$ is the viscosity of water, a good proxy for the viscosity of sweat, in $Pa\cdot s$ \citep{Kestin1978waterviscosity}, $L$ is the total length of the sweat gland, including both the secretory coil and the duct, in $m$ \citep{Wilke2007sweatgland}, and $d$ is the inner luminal diameter of the sweat gland in $m$ \citep{Hibbs1958}. The velocity of water flow in the sweat gland compartment, denoted as $u_{sg}$ in $m\,s^{-1}$, can be expressed as the ratio of the water flow rate inside the sweat gland $Q_{water,sg}$ to the inner luminal area of the sweat gland $A_{sg}$ and is given as

\begin{equation}
u_{sg} = \frac{Q_{water,sg}}{A_{sg}}. \label{equation5}
\end{equation}

At the same time, glucose is transported by the water through the sweat gland, eventually reaching the skin surface. This process can be described as a diffusion-convection process as
\begin{equation}
    \frac{\partial C_{sg,dil}}{\partial t}=D_{sw}\frac{\partial^2 C_{sg,dil}}{\partial y_{sg}^2}-u_{sg}\frac{\partial C_{sg,dil}}{\partial y_{sg}},\label{equ4}
\end{equation}
where $C_{sg,dil}$ represents the diluted glucose concentration in the sweat gland in $mol\ m^{-3}$, $u_{sg}$ is the flow velocity inside the sweat gland in $m\,s^{-1}$, and $D_{sw}$ is the diffusion coefficient of glucose in sweat, measured in $m^2\,s^{-1}$. Given that the sweat gland is primarily composed of water, $D_{sw}$ is assumed to be equivalent to the diffusion coefficient of glucose in water \citep{zhang2005effective21}. 

To account for the dilution effect of varying sweat rates on glucose concentrations, the diluted glucose concentration in the sweat gland $C_{sg,dil}$ can be given as
\begin{equation}
    C_{sg,dil}=C_{sg}\times\frac{1}{1+K_{w/g}\times u_{sweat,n}},\label{equ5}
\end{equation}
where $K_{w/g}$ is defined as the ratio of the volumetric flow rate of water to the volumetric flow rate of glucose \citep{li2022density22}, and $u_{sweat,n}$ represents the measured sweat velocity after normalization, both of which are dimensionless. This normalization is performed relative to the average passive sweat rate of $3\times10^{-4}$ $m\,s^{-1}$ \citep{Nie2018}. Normalizing the sweat rate ensures that variations in the ratio of the volumetric flow rate of water to glucose $K_{w/g}$ are meaningfully comparable across different sweating conditions.

All model parameters are sourced from existing literature and are listed in Table~\ref{t2}, and the full transport model was implemented using COMSOL Multiphysics\textsuperscript{\textregistered}. 

\begin{table*}
	
	\caption{Parameters of the Glucose Transport Model }
	\label{t2}
	\centering
	\scalebox{1}{
	\begin{tabular}{llll}
		\hline
		Parameter	&	Unit&	Value	&Ref   \\ 
		\hline
        
		Capillary hydrostatic pressure: $P_c$ & $mmHg$ & 30 & \cite{Haggerty2019} \\
		Capillary hydraulic conductivity of water: $L_{p,c}$ & $m\ s^{-1} \ mmHg^{-1}$ &	$6.5\times10^{-10}$ & \cite{Kellen2003} \\
        Diffusion coefficient of glucose for sweat gland wall: $D_{sg,wall}$ &$m^2\,s^{-1}$ &$6.46\times10^{-10}$ & \cite{zhang2005effective21} \\
		Diffusion coefficient of glucose in ISF: $D_{ISF}$ & $m^2\,s^{-1}$& $2.64\times10^{-10}$	&	\cite{Khalil2006}\\

        Diffusion coefficient of glucose in sweat: $D_{sw}$ & $m^2\,s^{-1}$& $6.7\times10^{-10}$	&	\cite{zhang2005effective21}\\

		Effective surface area of capillary: $A_c$& $m^2$ & $1.5\times10^{-8}$&\cite{Himeno2016}\\
		Effective cross-sectional area of ISF: $A_{ISF}$& $m^2$ & $2.2\times10^{-8}$&\cite{Himeno2016}\\
		
		Effective volume of capillary: $V_{p}$& $m^3$ & $3.02\times10^{-13}$&\cite{Himeno2016}\\
		Effective volume of ISF: $V_{ISF}$& $m^3$ & $6.0\times10^{-13}$&\cite{Himeno2016}\\
		Interstitial hydrostatic pressure: $P_{ISF}$ & $mmHg$ & -3 & \cite{Haggerty2019} \\
		Inner luminal diameter of sweat gland: $d$ & $m$ & $5\times10^{-6}$ & \cite{Hibbs1958}\\
		
		Glucose clearance constant: $k_{DE}$ & $s^{-1}$& $6.51\times10^{-4}$	&	\cite{Ibrahim2012}\\
		Glucose uptake rate: $r_{uptake}$	& $mol\ m^{-3}\,s^{-1}$  	&	$2.78\times10^{-2}$ & \cite{longo1992glucose17}\\
		Length of sweat gland: $L$ & $m$ & $4\times10^{-3}$ & \cite{Wilke2007sweatgland}\\

		Pressure difference across sweat gland: $\Delta P$ & $mmHg$ & 10 & \cite{Schulz1969} \\
		Ratio of volumetric flow rate of water to glucose: $K_{w/g}$ & - & 12 & \cite{li2022density22} \\
        Thickness of sweat gland wall: $h_{sg}$ & $m$ & $5\times10^{-5}$ & \cite{Sonner2015} \\
		Viscosity of water: $\mu$ & $Pa\cdot s$ & $1\times10^{-3}$ & \cite{Kestin1978waterviscosity} \\

		\hline\\
	\end{tabular}}
\end{table*}

\subsubsection{Model Performance Evaluation}
To characterize the temporal dynamics of our glucose transport model, we evaluated its step response, i.e., the variation in sweat glucose concentration resulting from a sudden variation in blood glucose concentration, formulated as a step function.

To assess the predictive performance of the model, we used seven sets of experimental data \citep{moyer2012correlation8,la2019modeling9,lee2016graphene15,boysen1984modified16,Nyein2019,Klous2021}. First, we compared the performance of our glucose transport model with that of La Count et al.'s work, the only other existing pharmacokinetic model that describes the kinetic transport mechanism of glucose from blood to sweat \citep{la2019modeling9}, focusing on root mean square error (RMSE) and root mean square percentage error (RMSPE) \citep{Taraji2017}. This comparison was carried out individually for each of the seven studies to assess the estimation results against the experimentally measured sweat glucose concentrations. Subsequently, we applied Wilcoxon signed-rank tests to the absolute error and absolute percentage error of the estimation results across all 108 measurements points to determine if there were significant differences in estimation performance between our model and La Count et al.'s work. Additionally, the Pearson correlation coefficient ($R$) was calculated for the aggregated 108 measurement points from all studies, as the number of data points in each individual study is insufficient to provide reliable estimates of $R$.

In the sixth and seventh studies \citep{Nyein2019,Klous2021}, blood and sweat glucose concentrations for each subject were measured multiple times under different conditions, with only one measurement taken per condition. This contrasts with other studies where a series of continuous glucose measurements were taken under a single condition for each subject. Accordingly, we treated these single measurements of glucose concentrations as constants for each condition, assuming no variation over time.

\subsubsection{Sensitivity Analysis}
\par To evaluate the impact of the model parameter values on the estimation performance of the glucose transport model, we conducted a sensitivity analysis focusing on the biophysical parameters listed in Table~\ref{t2}. A total of 100 samples for each parameter were generated using a random number generator, each following a Gaussian distribution centered around the respective literature value listed in Table~\ref{t2}, with a standard deviation of $10\%$. The model estimation results were obtained for each parameter value, and the corresponding coefficients of variation (CV) were determined, where CV is defined as the ratio of the standard deviation to the mean \citep{KM23}.

\subsection{Double-Loop Optimization Strategy}
In our study, we aim to estimate blood glucose concentrations using a double-loop optimization strategy that minimizes an error function based on the measured sweat glucose concentrations. The primary source of error in the optimization approach originates from the initially estimated blood glucose concentration, which may not accurately reflect individual physiological conditions. A first optimization loop addresses this by refining the estimated blood glucose concentration. Additionally, the biophysical parameters of the model (Table~\ref{t2}), derived from the literature and subject to inter-person variation, also contribute to the error. A second optimization loop enhances the overall estimation accuracy by updating the values of the glucose transport model parameters in Table~\ref{t2} for each experiment. These updated values establish a personalized glucose transport model. The two loops are intertwined to effectively minimize the estimation error. The details of the double-loop optimization strategy are reported in the following sections.

\subsubsection{Inverse Problem and Error Definition}
The proposed glucose transport model describes the transport of glucose from blood to sweat. However, due to the model complexity, solving the inverse problem, i.e., calculating blood glucose concentrations from sweat measurements, is not straightforward and requires the employment of optimization strategies. Therefore, we employ an optimization approach aimed at minimizing the error function between the sweat glucose concentrations estimated by our glucose transport model ($\hat{C}_{sweat\ glu,i}$) and those experimentally measured ($C_{sweat\ glu}$). The error function, denoted as $e_i$, is quantified as
\begin{equation}
    e_i=\frac{1}{N} \sum_{n=1}^N (C_{sweat\ glu,n} - \hat{C}_{sweat\ glu,n,i})^2, \label{equ6}
\end{equation}

where $C_{sweat\ glu,n}$ is the measured glucose concentration in sweat of the nth data point, $\hat{C}_{sweat\ glu,n,i}$ is the model-estimated glucose concentration in sweat for the same point, subscript $i$ denotes the iteration number of the optimization loop, and $N$ represents the number of data points in the experiment.

\begin{figure}[h]
    \centering
    \includegraphics[width=0.7\linewidth]{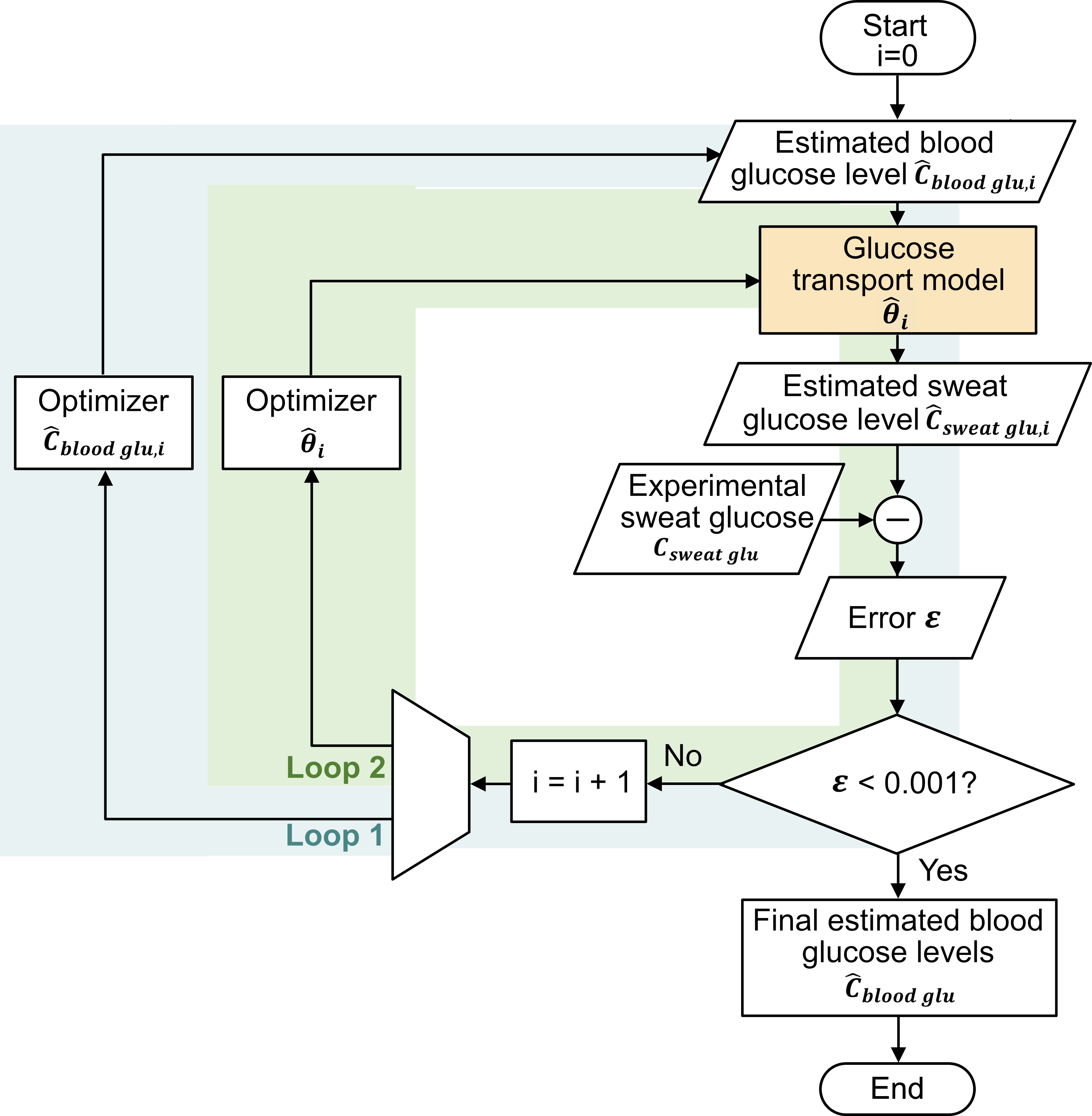}
    \caption{Flowchart of the double-loop optimization strategy. The intertwined optimization loops iteratively refining the estimated blood glucose concentrations (Loop 1) and the parameters of the glucose transport model (Loop 2), with the latter achieving a patient-specific approach.}
    \label{fig2}
\end{figure}

\subsubsection{Optimization Process}
Building on this, we introduce a double-loop optimization strategy, containing two distinct optimization loops as depicted in Fig.~\ref{fig2}. To initialize the optimization process, the initial estimate value for blood glucose concentration, denoted by $\hat{C}_{blood\ glu,0}$, is set to $5.5\ mmol\ L^{-1}$, based on the average values observed in healthy individuals as reported by \citep{danaei2011national24}. Table~\ref{t2} reports the initial values of the parameters of the glucose transport model, along with their corresponding literature sources. 

The first loop aims at adjusting the input value of blood glucose concentration, $\hat{C}_{blood\ glu,i}$, to minimize the error. Each single iteration of the first loop is followed by a single iteration of the second loop, which involves incremental adjustments to the person-specific biophysical parameters within the glucose transport model to further reduce the error. Although Table~\ref{t2} lists all the parameters of the glucose transport model, only those parameters with a corresponding coefficients of variation (CV) greater than 1\%, as identified through sensitivity analysis, are optimized in the second loop. This is because parameters with a CV less than 1\% are shown not to considerably impact the estimation performance of the model. By optimizing parameters with larger CVs, unnecessary computations were avoided, thereby improving the model's computational efficiency. This process of alternating single iterations between the first and second loops continues until the error, $e_i$, as defined in (\ref{equ6}), is reduced to less than 0.001 $mmol^2 \ L^{-2}$. Once this error threshold is met, the estimated glucose concentration in sweat closely aligns with the measured concentration, indicating that the input estimated glucose concentration in blood accurately reflects the actual blood glucose concentration. A sparse nonlinear optimizer was chosen for both loops due to its computational efficiency \citep{gill12saunders25}. To further enhance the robustness of the introduced approach, the estimation was performed over a sliding temporal window. Each window contained three experimental data points and moved with steps of one point. Consequently, each experimental data point underwent three estimations, and the mean of these values was adopted as the definitive estimated value.

\subsubsection{Model Performance Evaluation}
To evaluate the performance of our model, we conducted two sets of analyses. First, we compared the performance of two approaches: the single loop, which use only the first loop to optimize blood glucose concentrations without adjusting biophysical parameters, and the double-loop optimization strategy. This comparison was carried out individually for each of the seven studies, using RMSE and RMSPE to assess the estimation results against the experimentally measured blood glucose concentrations. Subsequently, Wilcoxon signed-rank tests were performed on the absolute error and absolute percentage error of the estimation results across a total of 108 experimental points from all studies, to evaluate whether there were statistically significant differences in estimation performance between the two models. Additionally, the Pearson correlation coefficient ($R$) was calculated for the aggregated 108 measurement points from all studies, as the number of data points in each individual study is insufficient to provide reliable estimates of $R$.

\begin{figure}[h]
    \centering
    \includegraphics[width=0.9\linewidth]{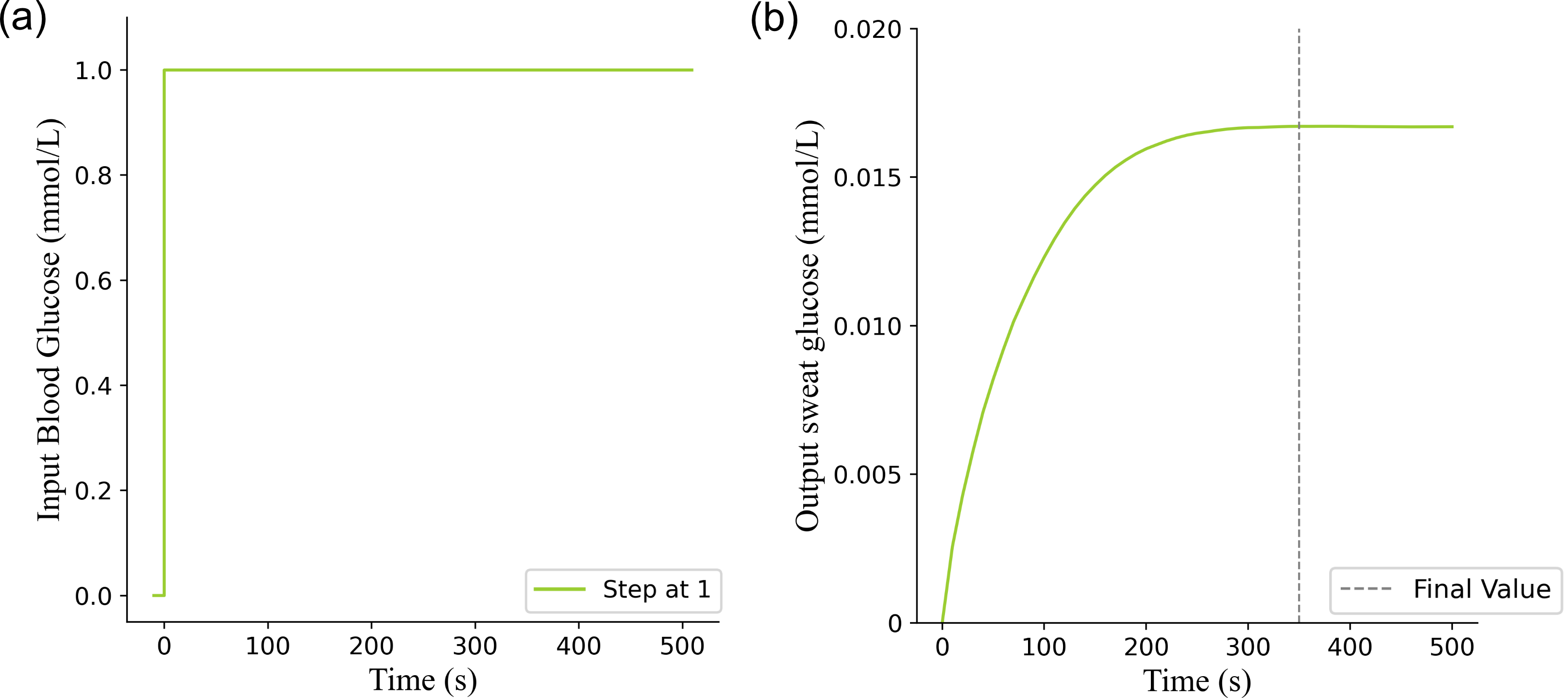}
    \caption{ Temporal dynamics of the glucose transport model to changes in glucose concentrations: (a) input blood glucose concentrations induced by a step function and (b) output sweat glucose concentration, with the dashed line indicating the time to steady states.}
    \label{fig3_step functions}
\end{figure}

\begin{figure*}[h]
    \centering
    \includegraphics[width=1\linewidth]{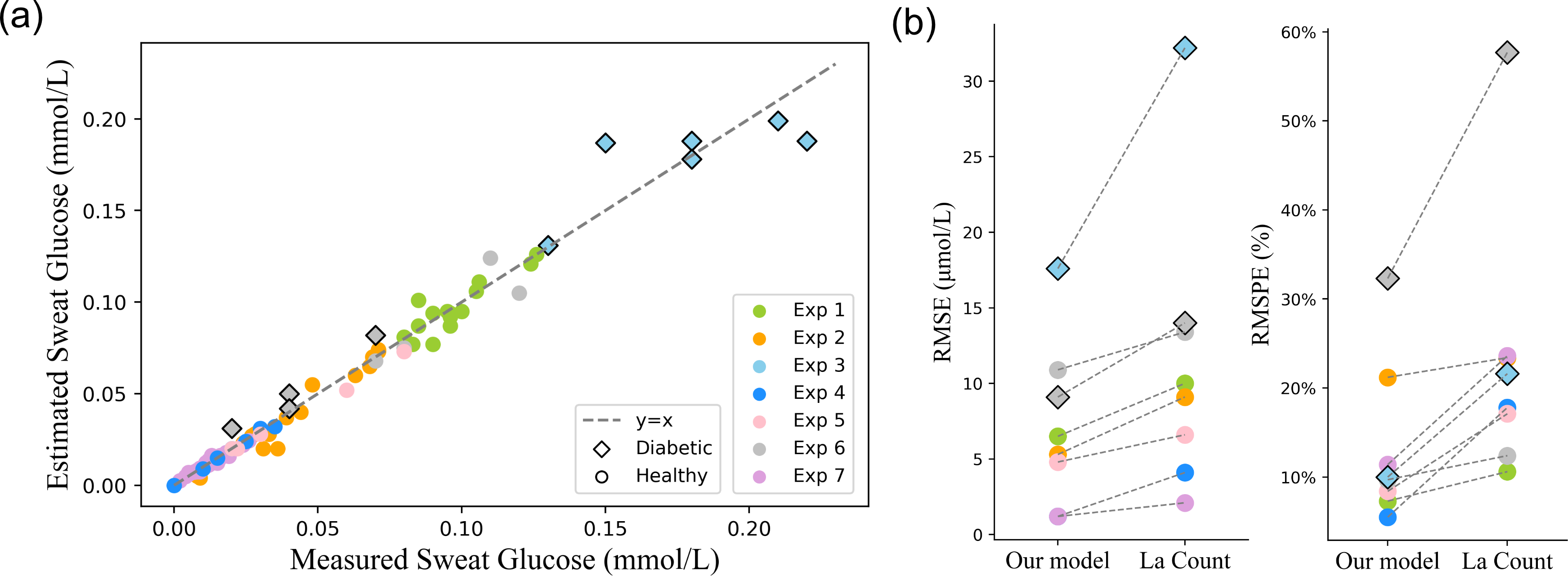}
    \caption{(a) Comparison of measured and estimated sweat glucose concentrations using our glucose transport model for the whole 108 experimental data points, where circles represent healthy subjects and outlined diamonds represent diabetic patients. See Table~\ref{t1} for an overview of the different experimental datasets. (b) Evaluation of our model and La Count et al.'s work by RMSE and RMSPE across seven studies. }
    \label{fig4_forward performance}
\end{figure*}

\begin{figure}[h]
    \centering
    \includegraphics[width=0.7\linewidth]{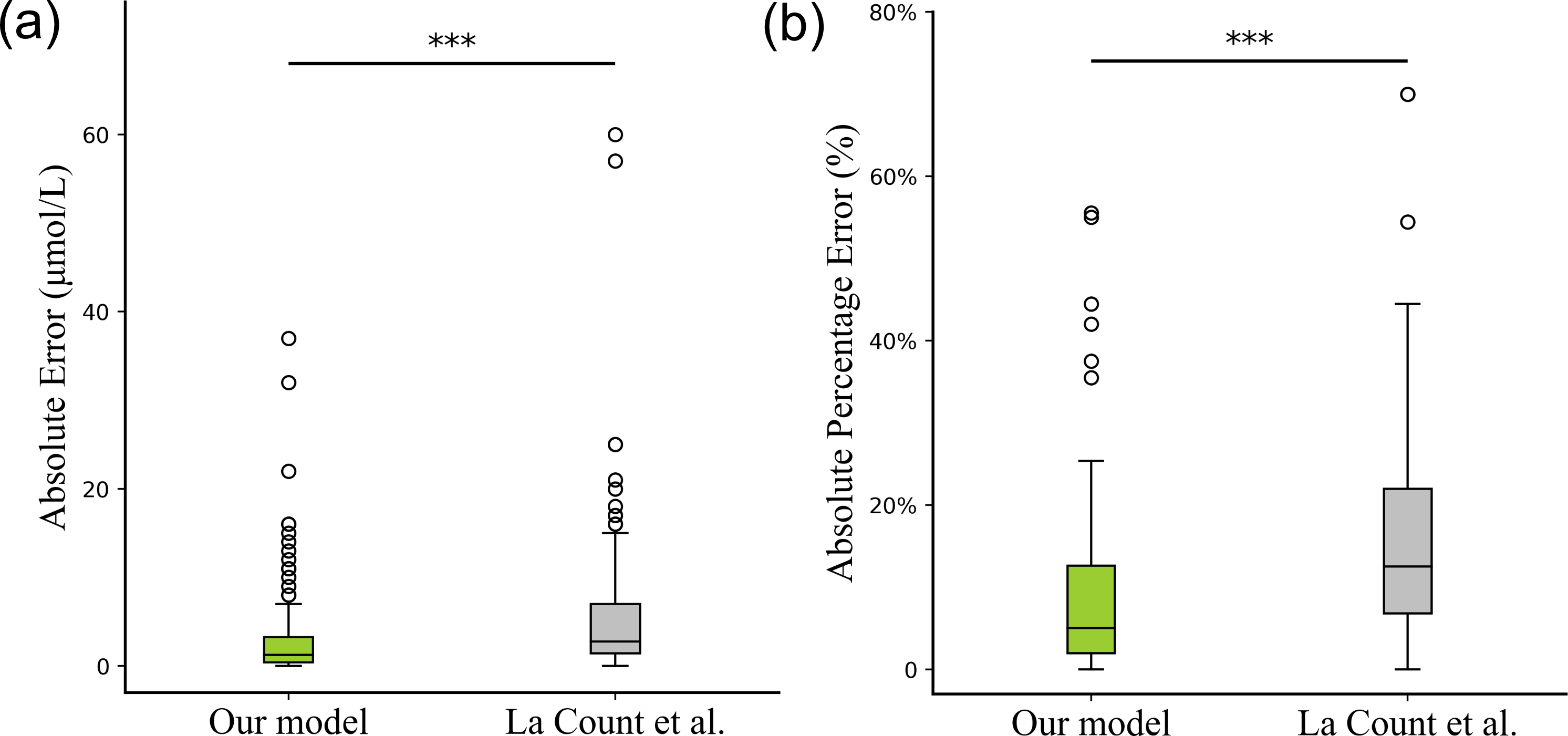}
    \caption{ Boxplot comparison of direct estimation performance by the glucose transport model: (a) absolute error and (b) absolute percentage error across all 108 experimental data points from the seven datasets. Asterisks denote statistical significance, `***' denotes statistical significance with $p<0.001$. Circles represent outliers. }
    \label{fig5_forward boxplot}
\end{figure}

\begin{table}[h]
    \centering
    \caption{Sensitivity Analysis of The Glucose Transport Model}
    \label{t3}
    \begin{threeparttable}
        \begin{tabular}{llc}
            \hline
            Parameter & Unit & Coefficient of variation (CV) \\ 
            \hline
            $D_{sg,wall}$ & $m^2\,s^{-1}$ & 3.6\% \\
            $D_{sw}$ & $m^2\,s^{-1}$ & 7.9\% \\
            $K_{w/g}$ & - & 18.2\% \\
            $h_{sg}$ & $m$ & 4.0\% \\
            \hline
        \end{tabular}
        \begin{tablenotes}
            \item Diffusion coefficient of glucose for sweat gland wall $D_{sg,wall}$, diffusion coefficient of glucose in sweat $D_{sw}$, ratio of volumetric flow rate of water to glucose $K_{w/g}$, and thickness of sweat gland wall $h_{sg}$.
        \end{tablenotes}
    \end{threeparttable}
\end{table}

\section{RESULTS}
\subsection{Glucose Transport Model}
The responsiveness of our glucose transport model to sudden changes in blood glucose concentrations is illustrated in Fig.~\ref{fig3_step functions}. The output sweat glucose concentrations reach their final values within approximately six minutes (350 seconds). This interval reflects the delay between an increase in glucose concentration in blood and the time it takes for an observable and stable glucose concentration increase in sweat.

The results of our glucose transport model are presented in Fig.~\ref{fig4_forward performance}(a), which translates experimentally measured glucose concentrations in blood into glucose concentrations in sweat. The estimation results of our model are also compared with those of La Count et al., as shown in Fig.~\ref{fig4_forward performance}(b). To allow a fair comparison, the biophysical parameters employed in our model were sourced from existing literature and were not optimized for individual patients.

Including the whole datasets with 108 experimental observations, the Pearson correlation coefficient $R$ for our model stands at 0.99, while for La Count et al.'s model, it is at 0.96. The average RMSE and RMSPE achieved by our model across the seven studies were 7$\pm$6 $\mu mol \,L^{-1}$ and 13$\%\pm$7$\%$. In comparison, the work of La Count et al. achieved average RMSE and RMSPE values of 11$\pm$10 $\mu mol \,L^{-1}$ and 22$\%\pm$10$\%$, respectively. The results of the Wilcoxon signed-rank tests, as shown in Fig.~\ref{fig5_forward boxplot}, revealed statistically significant differences between the estimation performance of our model compared to La Count et al.'s in both absolute error and absolute percentage error.

In the analysis of seven experimental datasets, the studies that included diabetic subjects, specifically the third and sixth datasets \citep{moyer2012correlation8,Nyein2019}, demonstrated the highest estimation errors as reflected by the corresponding RMSE and RMSPE values, respectively. The most substantial improvements using our model were observed in the third and sixth studies, with RMSPE reductions of 11.6\% and 17.9\%, respectively, when compared to the work of La Count et al. Additionally, there was a decrease in RMSE of 14.6 $\mu mol \,L^{-1}$ for the third study and 3.7 $\mu mol \,L^{-1}$ for the diabetic subset of the sixth study.

\par The results of the sensitivity analysis are presented in Table~\ref{t3}, which shows the parameters of the glucose transport model along with their respective CV values. Only parameters with a CV higher than 1\% are reported in the table. The ratio of flow rate of water to glucose $K_{w/g}$ results in the highest CV among these parameters, reaching 18.2$\%$, indicating that this parameter has the strongest influence on the estimation performance of the glucose transport model. The second highest CV is observed in the diffusion coefficient of glucose in sweat $D_{sw}$, with a CV of 7.9$\%$.

\begin{figure*}[h]
    \centering
    \includegraphics[width=1\linewidth]{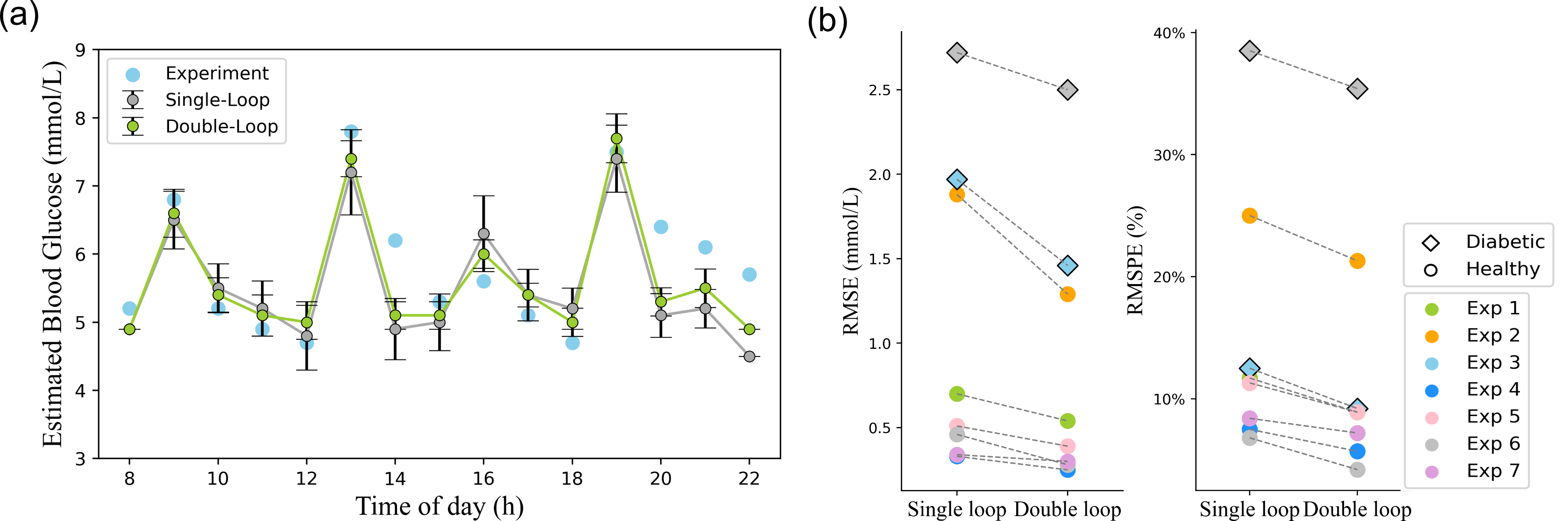}
    \caption{(a) Comparison of estimated blood glucose concentrations using the single-loop (represented in gray) and double-loop (represented in green) algorithms. Experimental data are represented by blue points, sourced from \citep{lee2016graphene15}. (b) Evaluation of performance for single- and double-loop optimization strategies, with RMSE and RMSPE noted separately for each study (see Table~\ref{t1}). }
    \label{fig6_seperate inverse result}
\end{figure*}

\begin{figure}[h]
    \centering
    \includegraphics[width=0.7\linewidth]{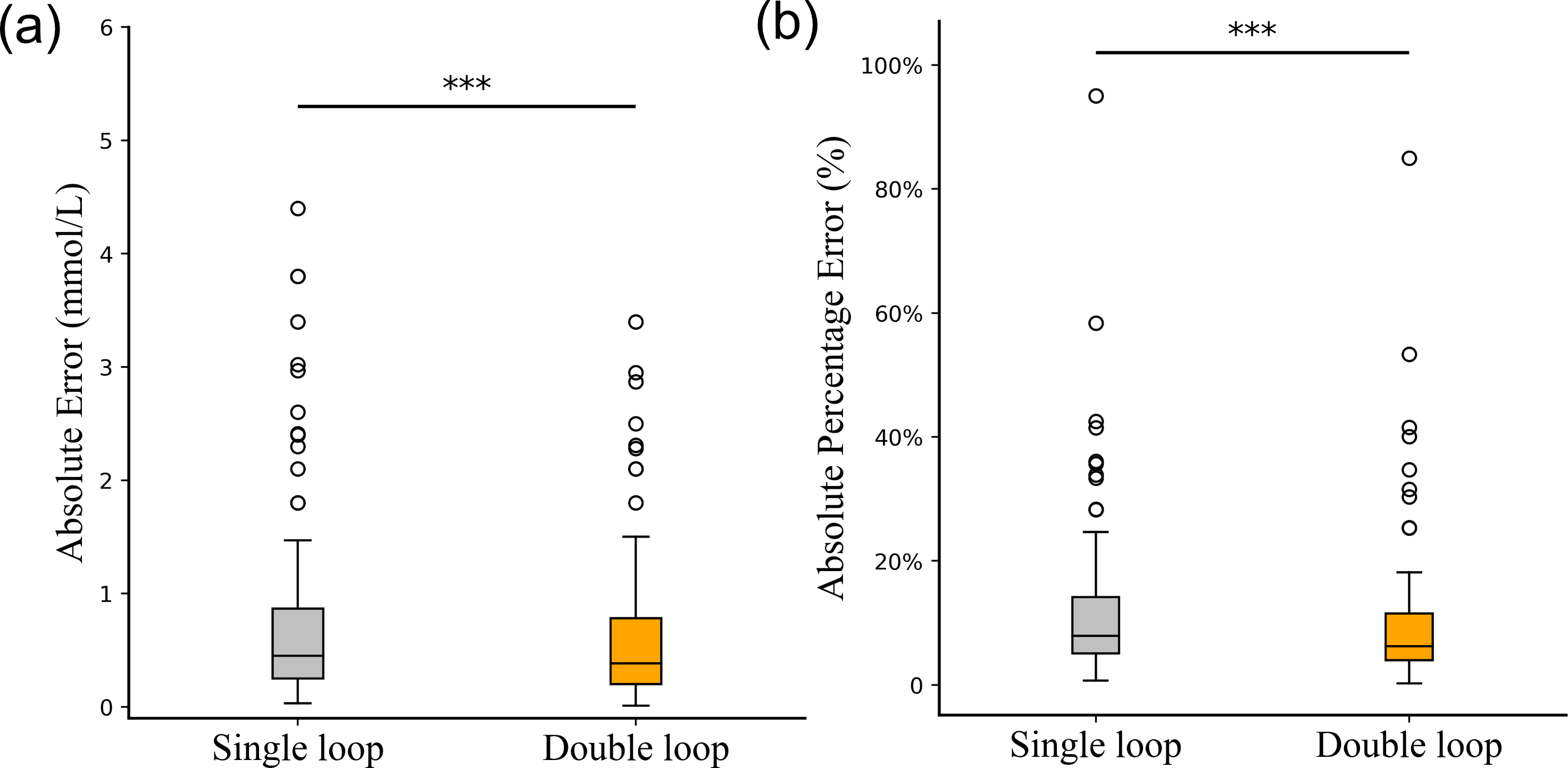}
    \caption{Boxplot comparison of inverse estimation performance of single-loop versus double-loop optimizations: (a) absolute error and (b) absolute percentage error across all 108 experimental data points from the seven datasets. Asterisks denote statistical significance, with `***' indicating $p<0.001$. Circles represent outliers.}
    \label{fig8_inverse boxplot}
\end{figure}

\begin{figure}[h]
    \centering
    \includegraphics[width=0.6\linewidth]{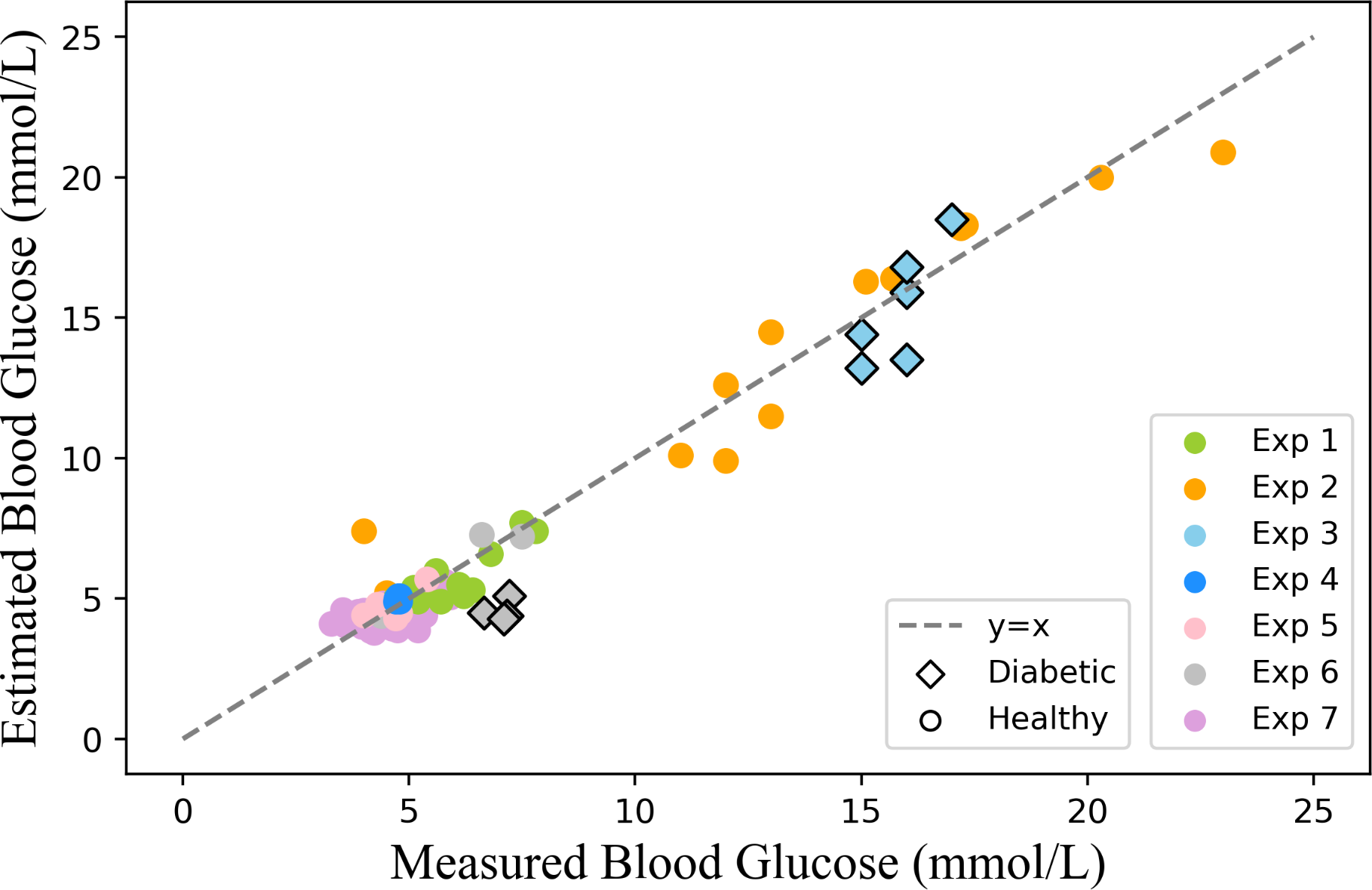}
    \caption{Comparison of estimated and measured blood glucose concentrations using double-loop optimization for the whole 108 experimental data points (see Table~\ref{t1}), where circles represent healthy subjects and outlined diamonds represent diabetic patients. }
    \label{fig7_inverse performance}
\end{figure}

\begin{table*}[h]
    \caption{Personalized Parameter Values of the Glucose Transport Model}
    \label{t4}
    \centering
    \begin{threeparttable}
        \begin{tabular}{llccc} 
            \hline
            Parameter & Initial Value & \multicolumn{3}{c}{Subject Health Condition} \\ 
            &  & \begin{tabular}[c]{@{}c@{}}Healthy \\(Min$\|$Max)\end{tabular} & \begin{tabular}[c]{@{}c@{}}Diabetic 1\\(Exp 3)\end{tabular} & \begin{tabular}[c]{@{}c@{}}Diabetic 2\\(Exp 6)\end{tabular} \\ 
            \hline
            $D_{sg,wall}$ ($m^2\,s^{-1}$) & $6.46\times10^{-10}$ & $(5.93\|8.62)\times10^{-10}$ & $1.30\times10^{-9}$ & $4.76\times10^{-10}$ \\
            $D_{sw}$ ($m^2\,s^{-1}$) & $6.7\times10^{-10}$ & $(5.66\|7.64)\times10^{-10}$ & $3.79\times10^{-10}$ & $7.82\times10^{-10}$ \\
            $K_{w/g}$ (-) & 12 & $(10.04\|13.36)$ & 10.40 & 12.82 \\
            $h_{sg}$ ($m$) & $5\times10^{-5}$ & $(3.45\|6.64)\times10^{-5}$ & $2.60\times10^{-5}$ & $6.67\times10^{-5}$ \\
            \hline
        \end{tabular}
        
        \begin{tablenotes}
            \item Diffusion coefficient of glucose for sweat gland wall $D_{sg,wall}$, diffusion coefficient of glucose in sweat $D_{sw}$, ratio of volumetric flow rate of water to glucose $K_{w/g}$, and thickness of sweat gland wall $h_{sg}$.
        \end{tablenotes}
    \end{threeparttable}
\end{table*}

\subsection{Single-Loop and Double-Loop Optimization}
We first conducted a single-loop optimization, where only the first loop that estimates blood glucose concentration was applied (refer to Fig.~\ref{fig2}). This process derived the glucose concentration in blood from the measured concentration in sweat, without optimizing the person-specific biophysical parameters of the glucose transport model, which were fixed at the average values reported in the literature. Subsequently, we executed the double-loop optimization, which includes the optimization of the model parameters that showed a CV higher than 1\%, as reported in Table~\ref{t3}. By comparing the results from both the single- and double-loop optimization strategies, we can evaluate the importance of incorporating person-specific biophysical parameters into the model's estimation performance. 

Figure~\ref{fig6_seperate inverse result}(a) displays the estimated blood glucose concentrations over time from both the single- and double-loop optimization strategies, as applied to an illustrative study (experimental data from the first study \citep{lee2016graphene15}). For the single-loop optimization, the average metrics across the seven studies were recorded as RMSE: 1.1$\pm$0.8 $mmol\,L^{-1}$ and RMSPE: 15$\%\pm$8$\%$. The double-loop optimization across the seven studies yielded an average RMSE and RMSPE of 0.9$\pm$0.6 $mmol\,L^{-1}$, 12$\%\pm$8$\%$, respectively. The results of the Wilcoxon signed-rank tests, as shown in Fig.~\ref{fig8_inverse boxplot}, revealed significant differences between the single-loop and the double-loop optimizations in both absolute error and absolute percentage error.

Figure~\ref{fig7_inverse performance} compares the estimated blood glucose concentration values to the measured values for all experimental data points, utilizing double-loop optimization. As for the total of 108 experimental points across the seven studies, our analysis revealed $R$ = 0.96 for the single-loop optimization and $R$ = 0.98 for the double-loop optimization, with $R^{2}$ equal to 0.92 and 0.95, respectively.

Table~\ref{t4} presents the personalized biophysical parameters derived from our glucose transport model using the double-loop optimization strategy. These values demonstrate our model's capability to customize model parameters to match an individual's unique physiological traits across different studies and are within a physiological range.

\section{Discussion}
In this study, we introduced a glucose transport model that translates experimental blood glucose concentrations into glucose concentrations in sweat. Furthermore, double-loop optimization strategy was proposed for the inverse estimation of the blood glucose concentrations from measured sweat glucose concentrations leading to a personalized model. To the best of our knowledge, this method is unprecedented. Biomarker concentrations in blood are currently considered the standard in clinical practice. Therefore, our approach could pave the way for the acceptance of future sweat-sensing technologies in clinical settings. 

\subsection{Glucose Transport Model}
The delay in glucose transport from blood to sweat, as estimated by our glucose transport model, is approximately six minutes. This aligns closely with the magnitude of delay times reported in the literature, typically around eight minutes \citep{moyer2012correlation8}. This consistency not only supports the validity of our model but can also help clinicians understand the value of future sweat-sensing technology. The results of the two-tailed Wilcoxon signed-rank tests indicate that the estimation performance of our model is significantly higher than that of the work by La Count et al., who developed the only other existing pharmacokinetic model that describes the kinetic transport mechanism of glucose from blood to sweat, with a p-value below 0.001. This observation demonstrates that the developed model not only improves the accuracy of sweat glucose concentration estimation but also offers a more robust characterization of the relationship between sweat and blood glucose concentrations. Specifically, improvements are evident in both the fit, as indicated by reduced RMSE and RMSPE, and the predictive power, evidenced by an increased Pearson correlation coefficient. Furthermore, our model uses only 18 parameters, which is much fewer than the 27 parameters used in La Count et al.'s work, highlighting its efficiency.

\par In the sensitivity analysis, the volumetric flow rate of water to glucose $K_{w/g}$ appears to largely impact the estimation accuracy of the model ($CV=18.2\%$). This suggests that the inclusion in our model of the glucose dilution effect due to sweat-rate related water influx into the sweat gland coil is indeed relevant. Altogether, our results indicate that the glucose transport model is sensitive to fluctuations in its biophysical parameters, which supports our patient-specific inference approach based on double-loop optimization.

\subsection{Single-Loop Optimization for Inverse Estimation}
Our single-loop optimization method, applied to infer blood glucose concentrations with fixed model parameters, demonstrated high estimation accuracy across the analysis of 108 experimental data points derived from seven studies. This method achieved a $R^2$ value of 0.92 and a $R$ of 0.96, notably surpassing the results of previous studies that relied on linear regression for analysis. For instance, Moyer et al. reported a moderate $R$ of 0.75 across 115 experimental samples \citep{moyer2012correlation8}. Klous et al. achieved an $R$ value of 0.73 from 48 data points \citep{Klous2021}. Nyein et al. found a lower correlation coefficient of 0.3 with 48 data points \citep{Nyein2019}. And Karpova et al. reported a $R$ of 0.75 from 25 data points \citep{Karpova2019}. These results indicate that the single-loop optimization approach proposed in this study, which incorporates a pharmacokinetic model of glucose transport from blood to sweat, outperforms the traditional linear regression analyses employed in the aforementioned studies in estimating blood glucose concentrations.

Moreover, it should be emphasized that the RMSE values obtained through inverse estimation using both single- and double-loop optimization strategies (Fig.~\ref{fig6_seperate inverse result}(b)) are higher than the RMSE values achieved with the glucose transport model (Fig.~\ref{fig4_forward performance}), which calculates the glucose concentration in sweat. This can be ascribed to the considerable differences in glucose concentration concentrations, with blood glucose concentration being approximately 50 to 100 times higher than that of glucose concentration in sweat \citep{moyer2012correlation8}.

\subsection{Double-Loop Optimization for Patient-Specific Estimation}
\par The double-loop optimization exhibited superior estimation performance compared to the single-loop approach in each study, as evidenced by its lower RMSPE (12$\%\pm$8$\%$), compared to the RMSPE of 15$\%\pm$8$\%$ for the single-loop method. A collective assessment using a Wilcoxon signed-rank test on all the 108 experimental data points demonstrates a statistically significant improvement in the estimation accuracy by the double-loop method over the single loop approach, with a p-value below 0.001. This can be attributed to the fact that the single-loop optimization method uses fixed values for the biophysical parameters of the glucose transport model, which are derived from literature studies \citep{longo1992glucose17,zhang2005effective21,Ibrahim2012,li2022density22,Himeno2016}. However, using these fixed parameter values may not sufficiently capture the unique biophysical characteristics inherent to each individual, which are influenced by a range of factors, including but not limited to age, gender, race, and lifestyle choices. Each of these factors contributes distinctly to an individual's physiological profile. Additionally, inaccuracies in sweat rate measurements during experiments may influence the accuracy of glucose concentration estimates, as these measurements are applied to the proposed model.

\par The double-loop optimization strategy introduced in this study effectively addresses the aforementioned challenges. By refining the biophysical parameters of the glucose transport model via a second intertwined loop, the double-loop optimization strategy adapts the biophysical parameter values to each subject, including the parameter $K_{w/g}$ that directly impacts sweat rates. This enhances the estimation performance and offers a notable improvement over the single-loop method with fixed parameters, enabling blood glucose estimates to be customized and tailored to individual patient situations. In summary, our double-loop optimization strategy provides a practical tool to respond to changes in sweat rate caused by different environmental factors like temperature and humidity, as well as varying intensities of physical activity.

In the second and sixth studies \citep{boysen1984modified16,Nyein2019}, the single-loop method's limitations were evident, with RMSPE values being 1.7 times and 1.9 times above the average, respectively. This was primarily due to the considerable fluctuation in blood glucose concentrations during these studies, as well as the inherent variation characteristic of diabetic patients in the sixth study. Specifically, the subject in the second study exhibited blood glucose concentrations well above the normal threshold for healthy subjects, which is 10 $mmol\,L^{-1}$ \citep{care27}. The double-loop strategy, in contrast, significantly improved the overall estimation accuracy, with the most evident reductions in RMSPE observed in the second study, decreasing by 3.7$\%$. Similarly, for the diabetic subject of the sixth study, a decrease of 2.5$\%$ in RMSPE was notably observed. These results show the effectiveness of the double-loop optimization strategy in accurately estimating blood glucose concentrations in individuals and conditions deviating from the population average, outperforming the single-loop optimization through model personalization. Our results suggest that subjects with blood glucose concentrations significantly exceeding the healthy range tend to exhibit worse estimation outcomes. Nonetheless, this issue can be substantially mitigated through the implementation of a double-loop optimization strategy. 

The primary limitation of this study is in the relatively small size of the datasets, particularly the insufficient number of diabetic subjects. This deficiency highlights the need for additional data to validate our model and ensure the reliability of our findings in diabetic patients. 

Finally, future research could build on the proposed approach to broaden the scope of sweat monitoring by including other biomarkers, such as cortisol and lactate. This extension would expand the potential utility of our optimization strategy for non-invasive monitoring across a wider range of diseases via sweat-sensing technology. 

\section{Conclusion}
This study introduces an improved model of the glucose transport mechanism from blood to sweat. Furthermore, a double-loop optimization strategy is also proposed, allowing for the personalization of this glucose transport model and its use in solving the inverse estimation of blood glucose concentrations from measured sweat glucose. The obtained promising results position sweat-sensing as a feasible option for non-invasive monitoring of blood glucose concentrations in both healthy and diabetic populations. The potential of the proposed approach for semi-continuous glucose monitoring based on sweat-sensing technology may pave the way for more effective diabetes management.

\section*{CRediT authorship contribution statement}
\textbf{X. Yin:} Conceptualization, Methodology, Writing – original draft, Writing – review \& editing. \textbf{E. Peri:} Conceptualization, Supervision, Writing – review \& editing. \textbf{E. Pelssers:} Project administration, Supervision, Writing – review \& editing. \textbf{J. den Toonder:} Funding acquisition, Writing – review \& editing. \textbf{L. Klous:} Writing – review \& editing. \textbf{H. Daanen:} Writing – review \& editing. \textbf{M. Mischi:} Conceptualization, Funding acquisition, Supervision, Writing – review \& editing.

\section*{Declaration of competing interest}
The authors declare that there is no conflicts of interests.

		
\section*{Acknowledgment}
This work was supported in part by the Dutch Research Council (NWO) under Grant SEDAS 18271.

\bibliographystyle{manuscript}
\bibliography{manuscript}

\begin{thebibliography}{49}
\expandafter\ifx\csname natexlab\endcsname\relax\def\natexlab#1{#1}\fi
\providecommand{\url}[1]{\texttt{#1}}
\providecommand{\href}[2]{#2}
\providecommand{\path}[1]{#1}
\providecommand{\DOIprefix}{doi:}
\providecommand{\ArXivprefix}{arXiv:}
\providecommand{\URLprefix}{URL: }
\providecommand{\Pubmedprefix}{pmid:}
\providecommand{\doi}[1]{\href{http://dx.doi.org/#1}{\path{#1}}}
\providecommand{\Pubmed}[1]{\href{pmid:#1}{\path{#1}}}
\providecommand{\bibinfo}[2]{#2}
\ifx\xfnm\relax \def\xfnm[#1]{\unskip,\space#1}\fi
\bibitem[{Bloom et~al.(2018)Bloom, Chen, and McGovern}]{bloom2018carga1}
\bibinfo{author}{D.~Bloom}, \bibinfo{author}{S.~Chen}, \bibinfo{author}{M.~McGovern},
\newblock \bibinfo{title}{La carga econ{\'o}mica de las enfermedades no transmisibles y la enfermedad mental: Resultados para costa rica, jamaica y per{\'u}},
\newblock \bibinfo{journal}{Revista Panamericana de Salud P{\'u}blica} \bibinfo{volume}{42} (\bibinfo{year}{2018}). \bibinfo{note}{\url{https://doi.org/10.1016/j.eswa.2021.114747}}.
\bibitem[{Magliano and Boyko(2021)}]{DE2}
\bibinfo{author}{D.~Magliano}, \bibinfo{author}{E.~J. Boyko},
\newblock \bibinfo{title}{10th edition. brussels: International diabetes federation},
\newblock \bibinfo{journal}{IDF diabetes atlas}  (\bibinfo{year}{2021}). \bibinfo{note}{\url{https://diabetesatlas.org/idfawp/resource-files/2021/07/IDF_Atlas_10th_Edition_2021.pdf}}.
\bibitem[{Saeedi et~al.(2019)Saeedi, Petersohn, Salpea, Malanda, Karuranga et~al.}]{saeedi2019global3}
\bibinfo{author}{P.~Saeedi}, \bibinfo{author}{I.~Petersohn}, \bibinfo{author}{P.~Salpea}, \bibinfo{author}{B.~Malanda}, \bibinfo{author}{S.~Karuranga}, et~al.,
\newblock \bibinfo{title}{Global and regional diabetes prevalence estimates for 2019 and projections for 2030 and 2045: Results from the international diabetes federation diabetes atlas},
\newblock \bibinfo{journal}{Diabetes research and clinical practice} \bibinfo{volume}{157} (\bibinfo{year}{2019}) \bibinfo{pages}{107843}. \bibinfo{note}{\url{https://doi.org/10.1016/j.diabres.2019.107843}}.
\bibitem[{Hermanns et~al.(2022)Hermanns, Ehrmann, Shapira, Kulzer, Schmitt, and Laffel}]{hermanns2022coordination4}
\bibinfo{author}{N.~Hermanns}, \bibinfo{author}{D.~Ehrmann}, \bibinfo{author}{A.~Shapira}, \bibinfo{author}{B.~Kulzer}, \bibinfo{author}{A.~Schmitt}, \bibinfo{author}{L.~Laffel},
\newblock \bibinfo{title}{Coordination of glucose monitoring, self-care behaviour and mental health: achieving precision monitoring in diabetes},
\newblock \bibinfo{journal}{Diabetologia} \bibinfo{volume}{65} (\bibinfo{year}{2022}) \bibinfo{pages}{1883--1894}. \bibinfo{note}{\url{https://doi.org/10.1007/s00125-022-05685-7}}.
\bibitem[{Grant et~al.(2015)Grant, Donner, Fradkin, Hayes et~al.}]{grant2015standards5}
\bibinfo{author}{R.~Grant}, \bibinfo{author}{T.~Donner}, \bibinfo{author}{J.~Fradkin}, \bibinfo{author}{C.~Hayes}, et~al.,
\newblock \bibinfo{title}{Standards of medical care in diabetes—2015},
\newblock \bibinfo{journal}{Diabetes care} \bibinfo{volume}{38} (\bibinfo{year}{2015}) \bibinfo{pages}{S1--94}. \bibinfo{note}{\url{https://doi.org/10.2337/dc15-S003}}.
\bibitem[{Hoffman et~al.(2023)Hoffman, McKeage, Xu, Ruddy, Nielsen, and Taberner}]{hoffman2023minimally6}
\bibinfo{author}{M.~S. Hoffman}, \bibinfo{author}{J.~W. McKeage}, \bibinfo{author}{J.~Xu}, \bibinfo{author}{B.~P. Ruddy}, \bibinfo{author}{P.~M. Nielsen}, \bibinfo{author}{A.~J. Taberner},
\newblock \bibinfo{title}{Minimally invasive capillary blood sampling methods},
\newblock \bibinfo{journal}{Expert review of medical devices} \bibinfo{volume}{20} (\bibinfo{year}{2023}) \bibinfo{pages}{5--16}. \bibinfo{note}{\url{https://doi.org/10.1080/17434440.2023.2170783}}.
\bibitem[{Lu et~al.(2022)Lu, Zada, Yang et~al.}]{lu2022microneedle7}
\bibinfo{author}{H.~Lu}, \bibinfo{author}{S.~Zada}, \bibinfo{author}{L.~Yang}, et~al.,
\newblock \bibinfo{title}{Microneedle-based device for biological analysis},
\newblock \bibinfo{journal}{Frontiers in Bioengineering and Biotechnology} \bibinfo{volume}{10} (\bibinfo{year}{2022}) \bibinfo{pages}{851134}. \bibinfo{note}{\url{https://doi.org/10.3389/fbioe.2022.851134}}.
\bibitem[{Leelarathna et~al.(2022)Leelarathna, Evans, Neupane, Rayman et~al.}]{Leelarathna2022}
\bibinfo{author}{L.~Leelarathna}, \bibinfo{author}{M.~L. Evans}, \bibinfo{author}{S.~Neupane}, \bibinfo{author}{G.~Rayman}, et~al.,
\newblock \bibinfo{title}{Intermittently scanned continuous glucose monitoring for type 1 diabetes},
\newblock \bibinfo{journal}{N Engl J Med} \bibinfo{volume}{387} (\bibinfo{year}{2022}) \bibinfo{pages}{1477--1487}. \bibinfo{note}{\url{https://doi.org/10.1056/NEJMoa2205650}}.
\bibitem[{Hermans et~al.(2020)Hermans, Van Der~Velden, and Gawalko}]{Hermans2020}
\bibinfo{author}{A.~N.~L. Hermans}, \bibinfo{author}{R.~M.~J. Van Der~Velden}, \bibinfo{author}{M.~e.~a. Gawalko},
\newblock \bibinfo{title}{On‐demand mobile health infrastructures to allow comprehensive remote atrial fibrillation and risk factor management through teleconsultation},
\newblock \bibinfo{journal}{Clinical Cardiology} \bibinfo{volume}{43} (\bibinfo{year}{2020}) \bibinfo{pages}{1232--1239}. \bibinfo{note}{\url{https://doi.org/10.1002/clc.23469}}.
\bibitem[{Moonen et~al.(2020)Moonen, Haakma, Peri et~al.}]{Moonen2020}
\bibinfo{author}{E.~J. Moonen}, \bibinfo{author}{J.~R. Haakma}, \bibinfo{author}{E.~Peri}, et~al.,
\newblock \bibinfo{title}{Wearable sweat sensing for prolonged, semicontinuous, and nonobtrusive health monitoring},
\newblock \bibinfo{journal}{View} \bibinfo{volume}{1} (\bibinfo{year}{2020}) \bibinfo{pages}{20200077}. \bibinfo{note}{\url{https://doi.org/10.1002/VIW.20200077}}.
\bibitem[{Van~Enter and Von~Hauff(2018)}]{VanEnter2018}
\bibinfo{author}{B.~J. Van~Enter}, \bibinfo{author}{E.~Von~Hauff},
\newblock \bibinfo{title}{Challenges and perspectives in continuous glucose monitoring},
\newblock \bibinfo{journal}{Chemical Communications} \bibinfo{volume}{54} (\bibinfo{year}{2018}) \bibinfo{pages}{5032--5045}. \bibinfo{note}{\url{https://doi.org/10.1039/C8CC01678J}}.
\bibitem[{Davis et~al.(2024)Davis, Heikenfeld, Milla, and Javey}]{Davis2024}
\bibinfo{author}{N.~Davis}, \bibinfo{author}{J.~Heikenfeld}, \bibinfo{author}{C.~Milla}, \bibinfo{author}{A.~Javey},
\newblock \bibinfo{title}{The challenges and promise of sweat sensing},
\newblock \bibinfo{journal}{Nature Biotechnology}  (\bibinfo{year}{2024}). \bibinfo{note}{\url{https://doi.org/10.1038/s41587-023-02059-1}}.
\bibitem[{Moyer et~al.(2012)Moyer, Wilson, Finkelshtein, Wong, and Potts}]{moyer2012correlation8}
\bibinfo{author}{J.~Moyer}, \bibinfo{author}{D.~Wilson}, \bibinfo{author}{I.~Finkelshtein}, \bibinfo{author}{B.~Wong}, \bibinfo{author}{R.~Potts},
\newblock \bibinfo{title}{Correlation between sweat glucose and blood glucose in subjects with diabetes},
\newblock \bibinfo{journal}{Diabetes technology and therapeutics} \bibinfo{volume}{14} (\bibinfo{year}{2012}) \bibinfo{pages}{398--402}. \bibinfo{note}{\url{https://doi.org/10.1089/dia.2011.0262}}.
\bibitem[{Klous et~al.(2021)Klous, De~Ruiter, Scherrer, Gerrett, and Daanen}]{Klous2021}
\bibinfo{author}{L.~Klous}, \bibinfo{author}{C.~De~Ruiter}, \bibinfo{author}{S.~Scherrer}, \bibinfo{author}{N.~Gerrett}, \bibinfo{author}{H.~Daanen},
\newblock \bibinfo{title}{The (in)dependency of blood and sweat sodium, chloride, potassium, ammonia, lactate and glucose concentrations during submaximal exercise},
\newblock \bibinfo{journal}{Eur J Appl Physiol} \bibinfo{volume}{121} (\bibinfo{year}{2021}) \bibinfo{pages}{803–816}. \bibinfo{note}{\url{https://doi.org/10.1007/s00421-020-04562-8}}.
\bibitem[{Nyein et~al.(2019)Nyein, Bariya, Kivimäki et~al.}]{Nyein2019}
\bibinfo{author}{H.~Y.~Y. Nyein}, \bibinfo{author}{M.~Bariya}, \bibinfo{author}{L.~Kivimäki}, et~al.,
\newblock \bibinfo{title}{Regional and correlative sweat analysis using high-throughput microfluidic sensing patches toward decoding sweat},
\newblock \bibinfo{journal}{Sci. Adv.} \bibinfo{volume}{5} (\bibinfo{year}{2019}) \bibinfo{pages}{eaaw9906}. \bibinfo{note}{\url{https://doi.org/10.1126/sciadv.aaw9906}}.
\bibitem[{Karpova et~al.(2019)Karpova, Shcherbacheva, Galushin et~al.}]{Karpova2019}
\bibinfo{author}{E.~V. Karpova}, \bibinfo{author}{E.~V. Shcherbacheva}, \bibinfo{author}{A.~A. Galushin}, et~al.,
\newblock \bibinfo{title}{Noninvasive diabetes monitoring through continuous analysis of sweat using flow-through glucose biosensor},
\newblock \bibinfo{journal}{Anal. Chem.} \bibinfo{volume}{91} (\bibinfo{year}{2019}) \bibinfo{pages}{3778--3783}. \bibinfo{note}{\url{https://pubs.acs.org/doi/10.1021/acs.analchem.8b05928}}.
\bibitem[{Mwaurah et~al.(2024)Mwaurah, Vinoth, Nakagawa et~al.}]{Mwaurah2024}
\bibinfo{author}{M.~M. Mwaurah}, \bibinfo{author}{R.~Vinoth}, \bibinfo{author}{T.~Nakagawa}, et~al.,
\newblock \bibinfo{title}{A neckband-integrated soft microfluidic biosensor for sweat glucose monitoring},
\newblock \bibinfo{journal}{ACS Appl. Nano Mater.} \bibinfo{volume}{7} (\bibinfo{year}{2024}) \bibinfo{pages}{17017--17028}. \bibinfo{note}{\url{https://pubs.acs.org/doi/10.1021/acsanm.4c03164}}.
\bibitem[{Sempionatto et~al.(2021)Sempionatto, Moon, and Wang}]{Sempionatto2021}
\bibinfo{author}{J.~R. Sempionatto}, \bibinfo{author}{J.-M. Moon}, \bibinfo{author}{J.~Wang},
\newblock \bibinfo{title}{Touch-based fingertip blood-free reliable glucose monitoring: Personalized data processing for predicting blood glucose concentrations},
\newblock \bibinfo{journal}{ACS Sens.} \bibinfo{volume}{6} (\bibinfo{year}{2021}) \bibinfo{pages}{1875--1883}. \bibinfo{note}{\url{https://pubs.acs.org/doi/10.1021/acssensors.1c00139}}.
\bibitem[{La~Count et~al.(2019)La~Count, Jajack, Heikenfeld, and Kasting}]{la2019modeling9}
\bibinfo{author}{T.~D. La~Count}, \bibinfo{author}{A.~Jajack}, \bibinfo{author}{J.~Heikenfeld}, \bibinfo{author}{G.~B. Kasting},
\newblock \bibinfo{title}{Modeling glucose transport from systemic circulation to sweat},
\newblock \bibinfo{journal}{Journal of pharmaceutical sciences} \bibinfo{volume}{108} (\bibinfo{year}{2019}) \bibinfo{pages}{364--371}. \bibinfo{note}{\url{https://doi.org/10.1016/j.xphs.2018.09.026}}.
\bibitem[{Mari et~al.(2020)Mari, Tura, Grespan, and Bizzotto}]{mari2020mathematical10}
\bibinfo{author}{A.~Mari}, \bibinfo{author}{A.~Tura}, \bibinfo{author}{E.~Grespan}, \bibinfo{author}{R.~Bizzotto},
\newblock \bibinfo{title}{Mathematical modeling for the physiological and clinical investigation of glucose homeostasis and diabetes},
\newblock \bibinfo{journal}{Frontiers in Physiology} \bibinfo{volume}{11} (\bibinfo{year}{2020}) \bibinfo{pages}{575789}. \bibinfo{note}{\url{https://doi.org/10.3389/fphys.2020.575789}}.
\bibitem[{Longo et~al.(1989)Longo, Griffin, Shuster, Langley, and Elsas}]{longo1989increased11}
\bibinfo{author}{N.~Longo}, \bibinfo{author}{L.~D. Griffin}, \bibinfo{author}{R.~C. Shuster}, \bibinfo{author}{S.~Langley}, \bibinfo{author}{L.~J. Elsas},
\newblock \bibinfo{title}{Increased glucose transport by human fibroblasts with a heritable defect in insulin binding},
\newblock \bibinfo{journal}{Metabolism} \bibinfo{volume}{38} (\bibinfo{year}{1989}) \bibinfo{pages}{690--697}. \bibinfo{note}{\url{https://doi.org/10.1016/0026-0495(89)90109-1}}.
\bibitem[{Basu et~al.(2006)Basu, Dalla~Man, Campioni et~al.}]{basu2006effects12}
\bibinfo{author}{R.~Basu}, \bibinfo{author}{C.~Dalla~Man}, \bibinfo{author}{M.~Campioni}, et~al.,
\newblock \bibinfo{title}{Effects of age and sex on postprandial glucose metabolism: differences in glucose turnover, insulin secretion, insulin action, and hepatic insulin extraction},
\newblock \bibinfo{journal}{Diabetes} \bibinfo{volume}{55} (\bibinfo{year}{2006}) \bibinfo{pages}{2001--2014}. \bibinfo{note}{\url{https://doi.org/10.2337/db05-1692}}.
\bibitem[{Mendes-Soares et~al.(2019)Mendes-Soares, Raveh-Sadka, Azulay, Edens et~al.}]{mendes2019assessment13}
\bibinfo{author}{H.~Mendes-Soares}, \bibinfo{author}{T.~Raveh-Sadka}, \bibinfo{author}{S.~Azulay}, \bibinfo{author}{K.~Edens}, et~al.,
\newblock \bibinfo{title}{Assessment of a personalized approach to predicting postprandial glycemic responses to food among individuals without diabetes},
\newblock \bibinfo{journal}{JAMA network open} \bibinfo{volume}{2} (\bibinfo{year}{2019}) \bibinfo{pages}{e188102--e188102}. \bibinfo{note}{\url{https://doi.org/10.1001/jamanetworkopen.2018.8102}}.
\bibitem[{Zenker et~al.(2007)Zenker, Rubin, and Clermont}]{zenker2007inverse14}
\bibinfo{author}{S.~Zenker}, \bibinfo{author}{J.~Rubin}, \bibinfo{author}{G.~Clermont},
\newblock \bibinfo{title}{From inverse problems in mathematical physiology to quantitative differential diagnoses},
\newblock \bibinfo{journal}{PLoS computational biology} \bibinfo{volume}{3} (\bibinfo{year}{2007}) \bibinfo{pages}{e204}. \bibinfo{note}{\url{https://doi.org/10.1371/journal.pcbi.0030204}}.
\bibitem[{Yin et~al.(2023)Yin, Peri, Pelssers et~al.}]{yin2023estimation28}
\bibinfo{author}{X.~Yin}, \bibinfo{author}{E.~Peri}, \bibinfo{author}{E.~Pelssers}, et~al.,
\newblock \bibinfo{title}{Estimation of blood glucose levels by sweat sensing based on biophysical modeling of glucose transport},
\newblock \bibinfo{journal}{2023 IEEE International Symposium on Medical Measurements and Applications (MeMeA)}  (\bibinfo{year}{2023}) \bibinfo{pages}{1--5}. \bibinfo{note}{\url{https://doi.org/10.1109/MeMeA57477.2023.10171952}}.
\bibitem[{Lee et~al.(2016)Lee, Choi, Lee, Cho et~al.}]{lee2016graphene15}
\bibinfo{author}{H.~Lee}, \bibinfo{author}{T.~K. Choi}, \bibinfo{author}{Y.~B. Lee}, \bibinfo{author}{Cho}, et~al.,
\newblock \bibinfo{title}{A graphene-based electrochemical device with thermoresponsive microneedles for diabetes monitoring and therapy},
\newblock \bibinfo{journal}{Nature nanotechnology} \bibinfo{volume}{11} (\bibinfo{year}{2016}) \bibinfo{pages}{566--572}. \bibinfo{note}{\url{https://doi.org/10.1038/nnano.2016.38}}.
\bibitem[{Boysen et~al.(1984)Boysen, Yanagawa, Sato, and Sato}]{boysen1984modified16}
\bibinfo{author}{T.~C. Boysen}, \bibinfo{author}{S.~Yanagawa}, \bibinfo{author}{F.~Sato}, \bibinfo{author}{K.~Sato},
\newblock \bibinfo{title}{A modified anaerobic method of sweat collection},
\newblock \bibinfo{journal}{Journal of Applied Physiology} \bibinfo{volume}{56} (\bibinfo{year}{1984}) \bibinfo{pages}{1302--1307}. \bibinfo{note}{\url{https://doi.org/10.1152/jappl.1984.56.5.1302}}.
\bibitem[{Ibrahim et~al.(2012)Ibrahim, Nitsche, and Kasting}]{Ibrahim2012}
\bibinfo{author}{R.~Ibrahim}, \bibinfo{author}{J.~M. Nitsche}, \bibinfo{author}{G.~B. Kasting},
\newblock \bibinfo{title}{Dermal clearance model for epidermal bioavailability calculations},
\newblock \bibinfo{journal}{Reviews in Chemical Engineering} \bibinfo{volume}{101} (\bibinfo{year}{2012}) \bibinfo{pages}{2094--2108}. \bibinfo{note}{\url{https://doi.org/10.1002/jps.23106}}.
\bibitem[{Himeno et~al.(2016)}]{Himeno2016}
\bibinfo{author}{Y.~Himeno}, et~al.,
\newblock \bibinfo{title}{Mechanisms underlying the volume regulation of interstitial fluid by capillaries: a simulation study},
\newblock \bibinfo{journal}{Integrative Medicine Research} \bibinfo{volume}{5} (\bibinfo{year}{2016}) \bibinfo{pages}{11--21}. \bibinfo{note}{\url{https://doi.org/10.1016/j.imr.2015.12.006}}.
\bibitem[{Wilke et~al.(2007)Wilke, Martin, Terstegen, and Biel}]{Wilke2007sweatgland}
\bibinfo{author}{K.~Wilke}, \bibinfo{author}{A.~Martin}, \bibinfo{author}{L.~Terstegen}, \bibinfo{author}{S.~S. Biel},
\newblock \bibinfo{title}{A short history of sweat gland biology},
\newblock \bibinfo{journal}{Intern J of Cosmetic Sci} \bibinfo{volume}{29} (\bibinfo{year}{2007}) \bibinfo{pages}{169--179}. \bibinfo{note}{\url{https://doi.org/10.1111/j.1467-2494.2007.00387.x}}.
\bibitem[{Kellen and Bassingthwaighte(2003)}]{Kellen2003}
\bibinfo{author}{M.~R. Kellen}, \bibinfo{author}{J.~B. Bassingthwaighte},
\newblock \bibinfo{title}{Transient transcapillary exchange of water driven by osmotic forces in the heart},
\newblock \bibinfo{journal}{American Journal of Physiology-Heart and Circulatory Physiology} \bibinfo{volume}{285} (\bibinfo{year}{2003}) \bibinfo{pages}{H1317--H1331}. \bibinfo{note}{\url{https://doi.org/10.1152/ajpheart.00587.2002}}.
\bibitem[{Haggerty and Nirmalan(2019)}]{Haggerty2019}
\bibinfo{author}{A.~Haggerty}, \bibinfo{author}{M.~Nirmalan},
\newblock \bibinfo{title}{Capillary dynamics, interstitial fluid and the lymphatic system},
\newblock \bibinfo{journal}{Anaesthesia and Intensive Care Medicine} \bibinfo{volume}{20} (\bibinfo{year}{2019}) \bibinfo{pages}{182--189}. \bibinfo{note}{\url{https://doi.org/10.1016/j.mpaic.2012.11.016}}.
\bibitem[{Rippe and Haraldsson(1986)}]{Rippe1986}
\bibinfo{author}{B.~Rippe}, \bibinfo{author}{B.~Haraldsson},
\newblock \bibinfo{title}{Capillary permeability in rat hindquarters as determined by estimations of capillary reflection coefficients},
\newblock \bibinfo{journal}{Acta Physiologica Scandinavica} \bibinfo{volume}{127} (\bibinfo{year}{1986}) \bibinfo{pages}{289--303}. \bibinfo{note}{\url{https://doi.org/10.1111/j.1748-1716.1986.tb07908.x}}.
\bibitem[{Longo et~al.(1992)Longo, Griffin, Langley, and Elsas}]{longo1992glucose17}
\bibinfo{author}{N.~Longo}, \bibinfo{author}{L.~D. Griffin}, \bibinfo{author}{S.~D. Langley}, \bibinfo{author}{L.~J. Elsas},
\newblock \bibinfo{title}{Glucose transport by cultured human fibroblasts: regulation by phorbol esters and insulin},
\newblock \bibinfo{journal}{Biochimica et Biophysica Acta (BBA)-Biomembranes} \bibinfo{volume}{1104} (\bibinfo{year}{1992}) \bibinfo{pages}{24--30}. \bibinfo{note}{\url{https://doi.org/10.1016/0005-2736(92)90127-8}}.
\bibitem[{Khalil et~al.(2006)Khalil, Kretsos, and Kasting}]{Khalil2006}
\bibinfo{author}{E.~Khalil}, \bibinfo{author}{K.~Kretsos}, \bibinfo{author}{G.~B. Kasting},
\newblock \bibinfo{title}{Glucose partition coefficient and diffusivity in the lower skin layers},
\newblock \bibinfo{journal}{Pharm Res} \bibinfo{volume}{23} (\bibinfo{year}{2006}) \bibinfo{pages}{1227--1234}. \bibinfo{note}{\url{https://doi.org/10.1007/s11095-006-0141-9}}.
\bibitem[{Lavery et~al.(2018)Lavery, Oldham, and Ghisalberti}]{Lavery2001}
\bibinfo{author}{P.~S. Lavery}, \bibinfo{author}{C.~E. Oldham}, \bibinfo{author}{M.~Ghisalberti},
\newblock \bibinfo{title}{The use of fick's first law for predicting porewater nutrient fluxes under diffusive conditions},
\newblock \bibinfo{journal}{Hydrological Processes} \bibinfo{volume}{15} (\bibinfo{year}{2018}) \bibinfo{pages}{2435--2451}. \bibinfo{note}{\url{https://doi.org/10.1002/hyp.297}}.
\bibitem[{Zhang and Fang(2005)}]{zhang2005effective21}
\bibinfo{author}{T.~Zhang}, \bibinfo{author}{H.~Fang},
\newblock \bibinfo{title}{Effective diffusion coefficients of glucose in artificial biofilms},
\newblock \bibinfo{journal}{Environmental technology} \bibinfo{volume}{26} (\bibinfo{year}{2005}) \bibinfo{pages}{155--160}. \bibinfo{note}{\url{https://doi.org/10.1080/09593332608618574}}.
\bibitem[{Sonner et~al.(2015)Sonner, Wilder, Heikenfeld et~al.}]{Sonner2015}
\bibinfo{author}{Z.~Sonner}, \bibinfo{author}{E.~Wilder}, \bibinfo{author}{J.~Heikenfeld}, et~al.,
\newblock \bibinfo{title}{The microfluidics of the eccrine sweat gland, including biomarker partitioning, transport, and biosensing implications},
\newblock \bibinfo{journal}{Biomicrofluidics} \bibinfo{volume}{9} (\bibinfo{year}{2015}) \bibinfo{pages}{031301}. \bibinfo{note}{\url{https://doi.org/10.1063/1.4921039}}.
\bibitem[{Aronson(2022)}]{Aronson2022}
\bibinfo{author}{D.~Aronson},
\newblock \bibinfo{title}{The interstitial compartment as a therapeutic target in heart failure},
\newblock \bibinfo{journal}{Front. Cardiovasc. Med.} \bibinfo{volume}{9} (\bibinfo{year}{2022}) \bibinfo{pages}{933384}. \bibinfo{note}{\url{https://doi.org/10.3389/fcvm.2022.933384}}.
\bibitem[{Schulz(1969)}]{Schulz1969}
\bibinfo{author}{I.~J. Schulz},
\newblock \bibinfo{title}{Micropuncture studies of the sweat formation in cystic fibrosis patients},
\newblock \bibinfo{journal}{J. Clin. Invest} \bibinfo{volume}{48} (\bibinfo{year}{1969}) \bibinfo{pages}{1470--1477}. \bibinfo{note}{\url{https://doi.org/10.1172/JCI106113}}.
\bibitem[{Kestin et~al.(1978)Kestin, Sokolov, and Wakeham}]{Kestin1978waterviscosity}
\bibinfo{author}{J.~Kestin}, \bibinfo{author}{M.~Sokolov}, \bibinfo{author}{W.~A. Wakeham},
\newblock \bibinfo{title}{Viscosity of liquid water in the range $-5$ to 150 $^\circ$c},
\newblock \bibinfo{journal}{J. Phys. Chem. Ref. Data} \bibinfo{volume}{7} (\bibinfo{year}{1978}) \bibinfo{pages}{941--948}. \bibinfo{note}{\url{https://doi.org/10.1063/1.555581}}.
\bibitem[{Hibbs(1958)}]{Hibbs1958}
\bibinfo{author}{R.~G. Hibbs},
\newblock \bibinfo{title}{The fine structure of human eccrine sweat glands},
\newblock \bibinfo{journal}{Am. J. Anat.} \bibinfo{volume}{103} (\bibinfo{year}{1958}) \bibinfo{pages}{201--217}. \bibinfo{note}{\url{https://doi.org/10.1002/aja.1001030204}}.
\bibitem[{Li et~al.(2022)Li, Wang, Yao, Cui, Gao, and Feng}]{li2022density22}
\bibinfo{author}{C.~Li}, \bibinfo{author}{N.~Wang}, \bibinfo{author}{Y.~Yao}, \bibinfo{author}{Z.~Cui}, \bibinfo{author}{J.~Gao}, \bibinfo{author}{D.~Feng},
\newblock \bibinfo{title}{Density, dynamic viscosity, conductivity and refractive index for mixture d-glucose and deep eutectic solvent (choline chloride+ urea) at different temperatures},
\newblock \bibinfo{journal}{Physics and Chemistry of Liquids} \bibinfo{volume}{60} (\bibinfo{year}{2022}) \bibinfo{pages}{83--94}. \bibinfo{note}{\url{https://doi.org/10.1080/00319104.2021.1916930}}.
\bibitem[{Nie et~al.(2018)Nie, Zhang, and Song}]{Nie2018}
\bibinfo{author}{S.~Nie}, \bibinfo{author}{C.~Zhang}, \bibinfo{author}{J.~Song},
\newblock \bibinfo{title}{Thermal management of epidermal electronic devices/skin system considering insensible sweating},
\newblock \bibinfo{journal}{Sci Rep} \bibinfo{volume}{8} (\bibinfo{year}{2018}) \bibinfo{pages}{14121}. \bibinfo{note}{\url{https://doi.org/10.1038/s41598-018-32152-4}}.
\bibitem[{Taraji et~al.(2017)Taraji, Haddad, Amos, Talebi, and Szucs}]{Taraji2017}
\bibinfo{author}{M.~Taraji}, \bibinfo{author}{P.~R. Haddad}, \bibinfo{author}{R.~I. Amos}, \bibinfo{author}{M.~Talebi}, \bibinfo{author}{R.~Szucs},
\newblock \bibinfo{title}{Error measures in quantitative structure-retention relationships studies},
\newblock \bibinfo{journal}{Journal of Chromatography A} \bibinfo{volume}{1524} (\bibinfo{year}{2017}) \bibinfo{pages}{298--302}. \bibinfo{note}{\url{https://doi.org/10.1016/j.chroma.2017.09.050}}.
\bibitem[{Krishnamoorthy and Lee(2024)}]{KM23}
\bibinfo{author}{K.~Krishnamoorthy}, \bibinfo{author}{M.~Lee},
\newblock \bibinfo{title}{Improved tests for the equality of normal coefficients of variation},
\newblock \bibinfo{journal}{Comput Stat} \bibinfo{volume}{29} (\bibinfo{year}{2024}) \bibinfo{pages}{215–232}. \bibinfo{note}{\url{https://doi.org/10.1007/s00180-013-0445-2}}.
\bibitem[{Danaei et~al.(2011)Danaei, Finucane, Lu, Singh et~al.}]{danaei2011national24}
\bibinfo{author}{G.~Danaei}, \bibinfo{author}{M.~M. Finucane}, \bibinfo{author}{Y.~Lu}, \bibinfo{author}{G.~M. Singh}, et~al.,
\newblock \bibinfo{title}{National, regional, and global trends in fasting plasma glucose and diabetes prevalence since 1980: systematic analysis of health examination surveys and epidemiological studies with 370 country-years and 2.7 million participants},
\newblock \bibinfo{journal}{The lancet} \bibinfo{volume}{378} (\bibinfo{year}{2011}) \bibinfo{pages}{31--40}. \bibinfo{note}{\url{https://doi.org/10.1016/S0140-6736(11)60679-X}}.
\bibitem[{Gill et~al.(2005)Gill, Murray, and Saunders}]{gill12saunders25}
\bibinfo{author}{P.~E. Gill}, \bibinfo{author}{W.~Murray}, \bibinfo{author}{M.~A. Saunders},
\newblock \bibinfo{title}{Snopt: An sqp algorithm for large-scale constrained optimization},
\newblock \bibinfo{journal}{SIAM Rev.} \bibinfo{volume}{47} (\bibinfo{year}{2005}) \bibinfo{pages}{99--131}. \bibinfo{note}{\url{https://doi.org/10.1137/S1052623499350013}}.
\bibitem[{Care(2019)}]{care27}
\bibinfo{author}{D.~Care},
\newblock \bibinfo{title}{6. glycemic targets: standards of medical care in diabetes—2019},
\newblock \bibinfo{journal}{Diabetes Care} \bibinfo{volume}{42} (\bibinfo{year}{2019}) \bibinfo{pages}{S61--70}. \bibinfo{note}{\url{https://doi.org/10.2337/dc19-S006}}.

\end{thebibliography}
	
\end{document}